\documentclass[%
 reprint,
 amsmath,amssymb,
 aps,
]{revtex4-2}

\newcommand*\diff{\mathop{}\!\mathrm{d}}

\newcommand*\N{\mathbb N}
\newcommand*\Z{\mathbb Z}

\newcommand*\te[1]{\text{#1}}

\newcommand*\p[1]{\left(#1\right)}
\newcommand*\ps[1]{\left[#1\right]}

\newcommand*\f[2]{\frac{#1}{#2}}
\newcommand*\mat[2]{\left(\begin{array}{#1}#2\end{array}\right)}

\newcommand*\I{\te{i}}
\newcommand*\pd[3]{\frac{\partial^{#3} #1}{\partial {#2}^{#3}}}
\newcommand*\td[3]{\frac{d^{#3}#1}{d #2^{#3}}}

\usepackage{color} 
\usepackage{hyperref}
\hypersetup{
    colorlinks=true, 
    linktoc=all, 
    linkcolor=black, 
}

\usepackage{float}
\usepackage{xcolor}
\usepackage{physics}
\usepackage{graphicx}
\usepackage{dcolumn}
\usepackage{bm}
\usepackage{fontawesome}
\usepackage{subfloat}

\usepackage{amsthm}

\theoremstyle{definition}

\begin{document}

\preprint{APS/123-QED}

\title{The Structure of the Oscillon\\\small{The dynamics of attractive self-interaction}}
\author{David Cyncynates}
 \email{davidcyn@stanford.edu}
\author{Tudor Giurgica-Tiron}%
 \email{tgt@stanford.edu}
\affiliation{%
 Stanford Institute for Theoretical Physics\\ Stanford University\\ 382 Via Pueblo, Stanford, CA 94305, USA
}

\date{\today}

\begin{abstract}
Real scalar fields with attractive self-interaction may form self-bound states, called oscillons. These dense objects are ubiquitous in leading theories of dark matter and inflation; of particular interest are long-lived oscillons which survive past $14\te{ Gyr}$, offering dramatic astrophysical signatures into the present day. We introduce a new formalism for computing the properties of oscillons with improved accuracy, which we apply to study the internal structure of oscillons and to identify the physical mechanisms responsible for oscillon longevity. In particular, we show how imposing realistic boundary conditions naturally selects a near-minimally radiating solution, and how oscillon longevity arises from its geometry. Further, we introduce a natural vocabulary for the issue of oscillon stability, which we use to predict new features in oscillon evolution. This framework allows for new efficient algorithms, which we use to address questions of whether and to what extent long-lived oscillons are fine-tuned. Finally, we construct a family of potentials supporting ultra-long-lived oscillons, with lifetimes in excess of $10^{17}$ years. \\
\faGithub\href{https://github.com/SimpleOscillon/Code}{\hspace{0.1cm}Public code.}
\end{abstract}

\maketitle

\tableofcontents

\section{\label{sec:MechanismsOfLifeAndDeath} Introduction}

Axions are real scalar fields predicted to exist in many extensions of the Standard Model.  One of the best-motivated is the QCD axion, which emerges as the pseudo-Nambu-Goldstone boson of a broken $U(1)$-axial symmetry, known as Peccei-Quinn (PQ) symmetry \cite{peccei1977constraints}. The PQ breaking scale $f_a$, known as the axion decay constant, suppresses the axion's self-interactions and its coupling to the Standard Model (SM). To avoid impacting stellar cooling rates, axion-SM interactions must be highly suppressed, forcing $f_a$ to be in the deep UV, $f_a\gtrsim 10^{10}\te{ GeV}$ \cite{raffelt2008astrophysical,anastassopoulos2017new,chang2018supernova}. As the universe cools below the QCD scale $\Lambda_\te{QCD}\approx 200\te{ MeV}$, strong dynamics generate a periodic potential for the axion, whose VEV cancels the strong sector's CP-violating phase, thus resolving the strong-CP problem. The separation between the QCD scale and the PQ scale forces the axion's mass $m_a\sim \Lambda_\te{QCD}^2/f_a$ to be smaller than $1$ meV, potentially by many orders of magnitude \cite{wilczek1978problem, weinberg1978new}.

Furthermore, axionic degrees of freedom emerge in great numbers from realistic string compactifications, collectively known as the Axiverse. Like the QCD axion, these axion-like particles (ALPs) are generally described by two parameters: their mass $m$ and the decay constant $f$. Generic ALPs are also expected to have naturally small masses, which are exponentially suppressed by the string instanton action. The precise form of the ALP potential depends on the specifics of the UV theory it descends from, leaving its low-energy dynamics effectively unconstrained \cite{arvanitaki2010string}.

\begin{subfigures}
\begin{figure}[t!]
    \centering
    \includegraphics[width = \columnwidth]{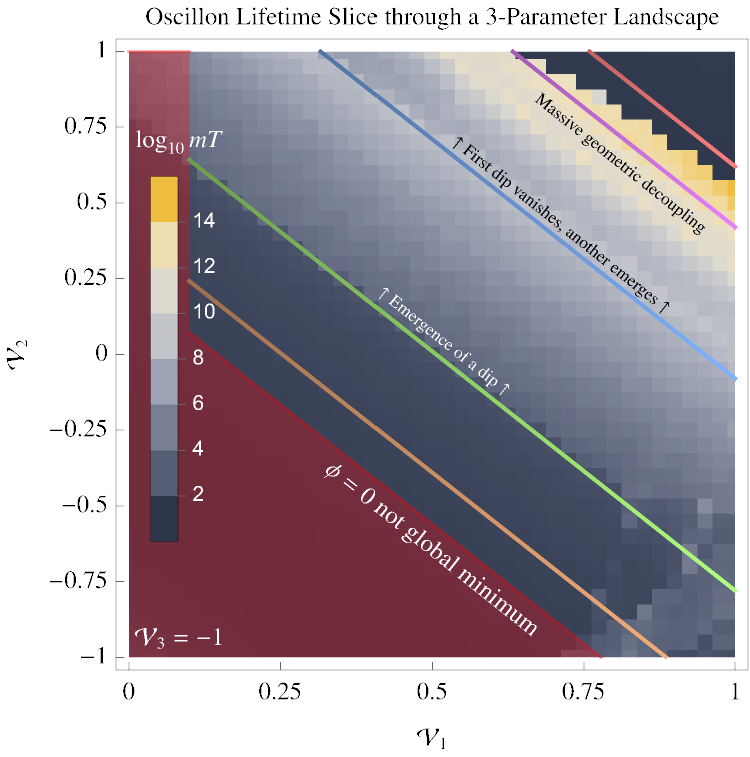}
    \caption{A slice through the oscillon lifetime landscape of parity symmetric periodic potentials with three free parameters \eqref{eq:PeriodicSum} (see text for details). The lifetime $T$ is calculated in units of the scalar mass $m$ for oscillons starting with a fundamental frequency of $\omega = 0.8 m$. The result is a glimpse into the structure of the oscillon lifetime landscape, revealing islands of longevity, separated by valleys. These features correspond to the location of exceptional `dip' frequencies, where the third harmonic experiences totally destructive interference. 
    We plot the families of potentials along the important colored contours in figure \ref{fig:PotentialSweep}.}
    \label{fig:landscape1}
\end{figure}
\begin{figure}
    \centering
    \includegraphics[width = \columnwidth]{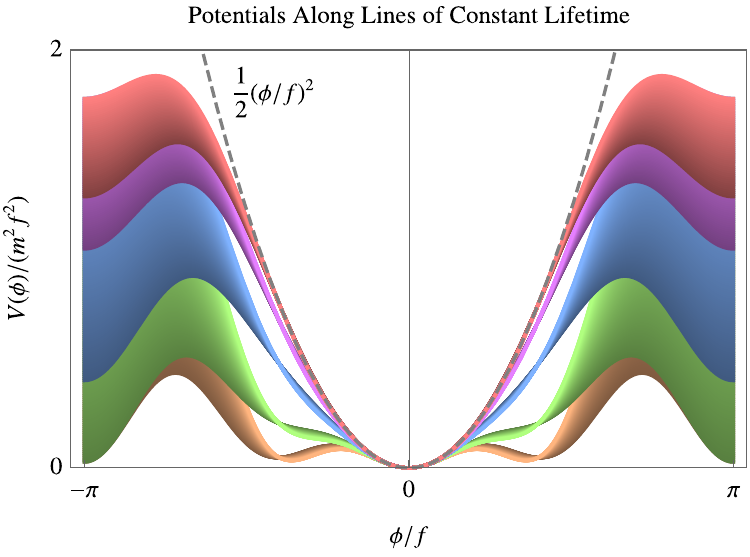}
    \caption{The potentials along the lines of constant lifetime in figure \ref{fig:landscape1}. To interpret this figure, we recognize that each color corresponds to approximately a single lifetime. Therefore, thin regions contain the most significant features, while broad regions, such as the value of the potential near $\phi/f = \pm\pi,$ are the least significant for determining the lifetime. As the central part of the potential approaches a free theory, the oscillon must grow in spatial extent because of weak self-interaction, leading to decoupling of the large bound oscillon from the short wavelength radiation (see section \ref{sec:geometricDeco}). On the other hand, some self-interaction is necessary to delay energetic death, which is why the purple potentials are much longer-lived than the red ones (see sections \ref{sec:EnergeticDeath} and \ref{sec:prescription}).}
    \label{fig:PotentialSweep}
\end{figure}
\end{subfigures}

Axions (both the QCD axion and ALPs) come equipped with natural production mechanisms, such as the vacuum misalignment mechanism, making them well-motivated dark matter candidates \cite{preskill1983cosmology,abbott1983cosmological,dine1983not,turner1990windows,sikivie2008axion,salvio2014thermal,di2016qcd,arvanitaki2020large,hall2020axion}. Of particular phenomenological interest are ultralight axions, whose masses can be as low as $10^{-21}\te{ eV}$ \cite{hui2017ultralight,irvsivc2017first,armengaud2017constraining,schutz2020subhalo,benito2020implications,hui2021wave,nadler2021dark,rogers2021strong}. Such ultralight axions lead to novel wave dark matter signatures, including effects on the matter power spectrum and structure formation \cite{du2016substructure,zhang2019cosmological,nori2019lyman,arvanitaki2020large,burkert2020fuzzy}, CMB observables \cite{fedderke2019axion,agrawal2020cmb}, and the formation of compact scalar structures such as axion minihalos \cite{kolb1993axion,buschmann2020early,blinov2020imprints}, gravitationally bound solitons and axion stars \cite{chavanis2018phase,braaten2018axion,visinelli2018dilute}, and self-interaction bound oscillons \cite{fodor2006oscillons,kudryavtsev1975solitonlike,makhankov1978dynamics,geicke1984cylindrical,gleiser1994pseudostable,kolb1994nonlinear,copeland1995oscillons,honda2002fine, saffin2007oscillons, fodor2008small, fodor2009radiation, gleiser2009general, amin2010flat, amin2012oscillons, andersen2012four, salmi2012radiation, mukaida2017longevity, fodor2019review, gleiser2019resonant,olle2020oscillons, zhang2020classical, olle2020recipes, kawasaki2020oscillon,hertzberg2010quantum,olle2019quantum,romanczukiewicz2018oscillons,dorey2020staccato}, the latter of which is the subject of this paper.

As the densest object in this family of bound axionic structures, oscillons promise dramatic astrophysical signatures, and have therefore been the subject of intense scrutiny \cite{van2018halometry,prabhu2020resonant,prabhu2020optical,kawasaki2020probing,amin2021electromagnetic,amin2021dipole,buckley2021fast}. Oscillons have a finite lifespan, and such phenomena crucially rely on oscillons that are cosmologically long-lived. Since dark matter axions are constrained by Lyman-$\alpha$ forest measurements to be at least $10^{-21}\te{ eV}$ in mass, their oscillation period is at most 0.1 years \cite{hui2017ultralight,irvsivc2017first,armengaud2017constraining,schutz2020subhalo,hui2021wave,nadler2021dark,rogers2021strong}. Therefore, oscillons that survive $14\te{ Gyr}$ until the present day must be stable for at least $10^{11}$ oscillations. Simulating an oscillon this long-lived is at the upper limit of current computational capabilities \cite{zhang2020classical,olle2020recipes}, and thus indirect methods are required to study longer-lived oscillons.

Significant progress has been made towards understanding the structure and evolution of oscillons in the last two decades, building on improved computational resources and theoretical understanding \cite{honda2002fine, saffin2007oscillons, fodor2008small, fodor2009radiation, gleiser2009general, hertzberg2010quantum, amin2010flat, amin2012oscillons, andersen2012four, salmi2012radiation, mukaida2017longevity, olle2019quantum, fodor2019review, gleiser2019resonant, zhang2020classical,olle2020oscillons, olle2020recipes, kawasaki2020oscillon,romanczukiewicz2018oscillons,dorey2020staccato}. Of central theoretical importance are artificial, exactly-periodic solutions of the equations of motion, which have been used to approximate the oscillon's instantaneous profile and radiation rate. In rare instances, in which the oscillon is known to be infinitely long-lived, this approximation is exact, and the solution is called a \emph{breather}: a finite energy periodic solution of the equations of motion. The most famous such example is the 1+1 dimensional sine-Gordon breather, which is stabilized by an infinite set of conserved quantities \cite{burt1978exact}. Breathers are not known to exist in 3+1 dimensions. Relaxing the breather's finite energy constraint, we find the periodic solutions known as \emph{quasibreathers}. These constructions have an infinite amount of energy residing in their standing-wave tails. These radiative tails can be understood as an approximation of the oscillon's classical radiation amplitude, which can be used to estimate the oscillon's lifetime.

In this paper, we further develop the quasibreather technique into a framework for understanding the classical properties of oscillons, unifying several observations made in the literature, and addressing key conceptual questions about the harmonic structure and stability of oscillons. By imposing realistic boundary conditions, we introduce the \emph{physical quasibreather} (PQB) as the member of the quasibreather family closest to a radiating oscillon, and arrive at an improved method for calculating oscillon properties, such as lifetime, radial profile, linear stability, and frequency content. In the limit of long lifetimes, our method becomes especially efficient, since semi-perturbative techniques may be employed to rapidly compute oscillon radiation. We apply our new methods to systematically study oscillon lifetimes in periodic axion potentials, allowing us to probe the genericity of long-lived oscillons. Moreover, we apply our framework to expand on existing studies of long-lived oscillons in monodromy potentials.

We summarize our study of oscillon lifetimes in periodic potentials with parity in the form of \emph{longevity landscapes}, such as the one depicted in figure \ref{fig:landscape1}. There, we scan the coefficients ${\mathcal V}_n$ of the axion potential $V(\phi)$ defined as
\begin{align}
\label{eq:PeriodicSum}
    V(\phi) = m^2 f^2\sum_{n = 1}^\infty \f{{\mathcal V}_n}{n^2}\p{1 - \cos\p{ \f{n\phi}{f}}}\,,\hspace{0.5cm} \sum_{n = 1}^\infty {\mathcal V}_n = 1\,.
\end{align}
Here, the field $\phi$ is the axion field, $m$ its mass, and $f$ its decay constant.
The particular slice through the space of coefficients in figure \ref{fig:landscape1} is defined by the choice to treat ${\mathcal V}_1$ and ${\mathcal V}_2$ as free parameters, while fixing ${\mathcal V}_3 = -1$, and forcing ${\mathcal V}_4$ to satisfy the mass constraint, with all other ${\mathcal V}_n$ set to zero. Our numerical techniques based on the PQB formalism have allowed us to perform this parameter sweep in 96 CPU-hours, parallelized down to a few hours of wall-clock time. We see that the landscape is broken down into ``islands of longevity,'' where neighboring potentials sustain oscillons that are similarly long-lived. 
While most of this space supports oscillons in the range $10^2 - 10^4$ oscillations, these few tunable parameters in the potential are enough to allow for oscillons that may live up to $10^{14}$ cycles.

\begin{figure}
    \centering
    \includegraphics[width = \columnwidth]{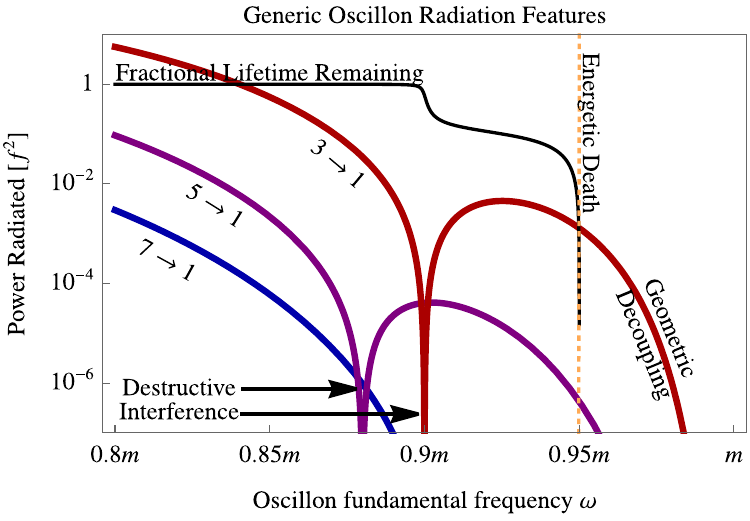}
    \caption{This plot illustrates the mechanisms of oscillon longevity and death described in section \ref{sec:lifecycle}. Here, we plot the power carried out of the oscillon in the dominant radiating harmonics as a function of the oscillon frequency $\omega$. The fundamental frequency $\omega$ increases with time, and therefore may be interpreted as a time coordinate (see figure \ref{fig:timeCurve}).
    For simplicity, we consider a scalar potential with parity symmetry, leading to radiation at odd multiples of $\omega$ due to $n\to 1$ processes.
    Towards higher frequencies, the size of the oscillon $2\pi/\sqrt{m^2 - \omega^2}$ is much larger than the radiation wavelength $2\pi/(n\omega)$, leading to the geometric decoupling of radiation. As the oscillon becomes more diffuse, its volume grows faster than its amplitude shrinks, forcing an early energetic death. 
    At exceptional frequencies, certain radiative harmonics vanish as a consequence of destructive self-interference.}
    \label{fig:genericPower}
\end{figure}

\begin{figure}
    \centering
    \includegraphics[width = \columnwidth]{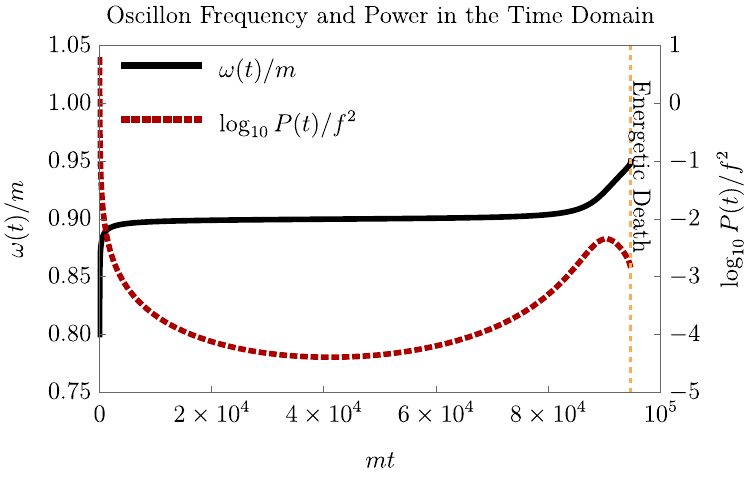}
    \caption{The oscillon's instantaneous frequency $\omega(t)$ and radiated power $P(t)$ plotted as explicit functions of time. These curves correspond to the generic scenario in figure \ref{fig:genericPower}. This plot illustrates how the oscillon spends most of its life at the exceptional frequency where the dominant radiating harmonic vanishes through destructive self-interference.
    }
    \label{fig:timeCurve}
\end{figure}

The distinct islands in figure \ref{fig:landscape1} correspond to the action of two mechanisms that suppress oscillon radiation,  which we identify as \emph{totally destructive self-interference} and \emph{geometric decoupling}. Together, these two effects comprise the \emph{form-factor} of the oscillon coupling to radiation, but we separate them because of their distinct imprints on the oscillon life-cycle, as depicted in figure \ref{fig:genericPower}. Further, the cliffs in figure \ref{fig:landscape1} represent destructive interference peaks entering unreachable frequencies beyond the point of \emph{energetic death}, where the oscillon is forced to dissipate because of energy conservation. Here, we briefly review these three effects. \\\\
\emph{Destructive interference:} The bound bulk of the oscillon is a nearly coherent object, oscillating at frequency $\omega$. Through the interaction terms of integer order $\phi^{n+1}$, the oscillon bulk behaves as a nearly coherent source of radiation at multiples of the fundamental frequency $n\omega$. Similar to a diffraction experiment, certain geometries lead to totally destructive interference, exponentially confining certain radiation channels at exceptional frequencies. When the dominant radiation channel destructively interferes, the radiated power experiences a sudden `dip.' \\\\
\emph{Geometric decoupling:} The size of the oscillon is inversely proportional to the binding energy per particle $\sqrt{m^2 - \omega^2}$, which blows up as $\omega$ approaches the rest mass $m$ (see figure \ref{fig:timeCurve}). In this limit, the oscillon grows much larger than the wavelengths of radiation $2\pi/n\omega$, causing a separation of scales. As this separation grows, the smooth oscillon bulk decouples from radiation, which manifests as an exponential decrease in radiated power towards the end of the oscillon's lifetime.\\\\
\emph{Energetic death:} As the oscillon radiates away its energy, the binding energy per particle decreases, reducing the oscillon's central amplitude and increasing its radius. In three or more spatial dimensions, weak self-interactions result in a volume growing faster than can be accommodated by the decreasing central amplitude. Therefore, at frequencies $\omega$ approaching the mass $m$, there is a point past which an external energy source is necessary for the oscillon to remain bound. At this point, the oscillon is forced to undergo a rapid process of dissipation, which we call energetic death.\\\\
\indent
These mechanisms explain the structure of the longevity landscape observed in figure \ref{fig:landscape1}. An island of longevity starts when a point of destructive interference (a `dip') emerges from low frequencies (green contour). As the dip migrates toward higher frequencies, its effect is enhanced by geometric decoupling, causing lifetime to increase until the dip moves beyond the frequency of energetic death, resulting in a longevity 'cliff' (blue contour).

In order to obtain these results, we have applied our PQB formalism to estimate the evolution of extremely long-lived spherically symmetric oscillons in isolation. In doing so, we have made the implicit assumption that the physical oscillon has relaxed into a state near the PQB. In order to check that this assumption is valid, we have performed a detailed linear stability analysis of the PQB to spherical and non-spherical perturbations, and we have presented evidence that unstable modes remain small enough that our procedure stays predictive (appendix \ref{sec:FloquetAnalysis}).

This paper is structured as follows. Section \ref{sec:physicalQuasibreatherMinTech} introduces the main object of study, the physical quasibreather. The oscillon is identified as living in the basin of attraction of the PQB, which naturally captures notions of oscillon stability. Section \ref{sec:lifecycle} uses the PQB formalism to understand the mechanisms of longevity briefly discussed above, and derives the minimum radiation condition. Section \ref{sec:prescription} applies the mechanisms of oscillon longevity and death to construct a family of potentials supporting ultra-long-lived oscillons. Section \ref{sec:tuned?} applies our techniques to study the genericity of long-lived oscillons, and introduces local and global measures of fine-tuning. Section \ref{sec:illustrativeExamples} applies our formalism to well-known potentials in the literature, re-deriving and expanding on previous results. Finally, the appendices provide a detailed technical overview of our formalism, and contain an exhaustive treatment of linear stability, as well as our numerical workflow. Appendix \ref{sec:ThePhysicalQuasibreatherFormalism} provides the mathematical basis of the PQB. Appendix \ref{sec:NM} details the numerical procedure for obtaining the PQB. Appendix \ref{sec:FloquetAnalysis} details our linear and nonlinear stability analysis of the PQB. Appendix \ref{sec:TechnicalFormulae} provides technical formulae relevant for computing the PQB and its linear stability. Appendix \ref{sec:simulation} details our explicit numerical simulations.

\section{\label{sec:physicalQuasibreatherMinTech}The physical quasibreather}
\begin{figure}
    \centering
    \includegraphics[width = \columnwidth]{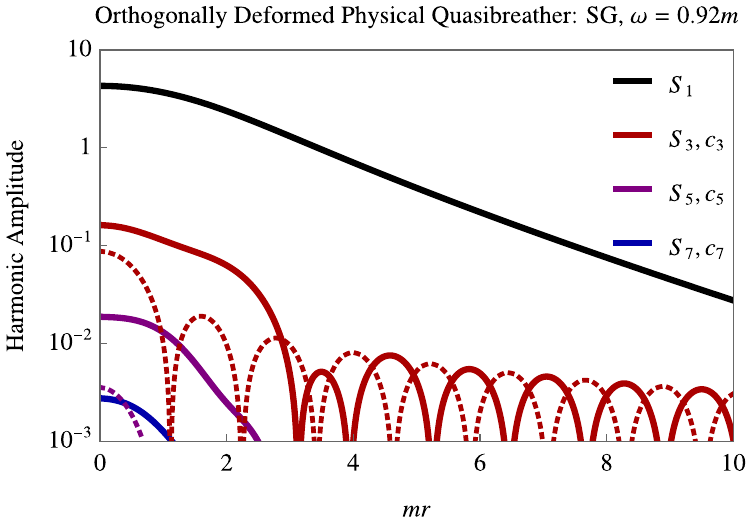}
    \caption{The radial profile of the physical quasibreather (PQB) (solid) and its orthogonal deformation (OD) (dashed) for the sine-Gordon (SG) oscillon at $\omega = 0.92 m$, plotted against radius in units of the mass $m^{-1}$. In the limit where the radiation tails are small, this serves as an instantaneous approximation of the internal structure of the oscillon. The first quasibreather harmonic $S_1$ is exponentially bound, defining the oscillon bulk. The third harmonic $S_3$ is the dominant radiation mode, followed by the fifth, seventh, and so on. The spatial and temporal phase of the OD are 90 degrees out of phase with the PQB in the radiative region, representing outgoing radiation.}
    \label{fig:DPQBAnatomy}
\end{figure}

\begin{figure}
    \centering
    \includegraphics[width = \columnwidth]{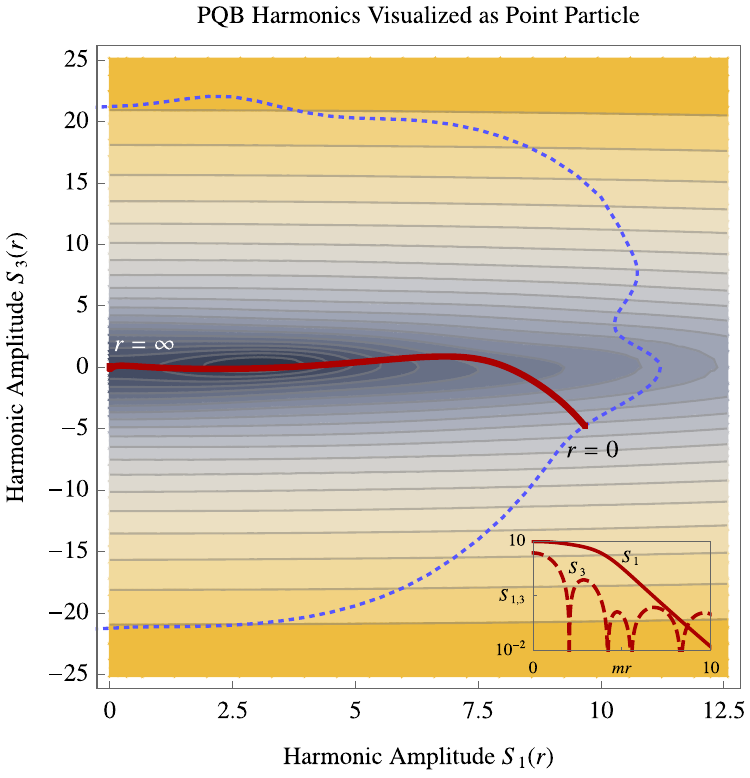}
    \caption{The PQB trajectory of the harmonic amplitudes $S_1$ and $S_3$ (red) is plotted on top of the level sets of the effective potential. The set of all initial conditions corresponding to quasibreathers is outlined in dotted blue. The particular example plotted here is of the sine-Gordon equation for $\omega = 0.5 m$.}
    \label{fig:rolling}
\end{figure}

The nonlinear wave equation we study in this paper is of the generic form
\begin{align}
\label{eq:EOM}
    0&=\ddot\phi - \nabla^2\phi + V'(\phi/f)\,.
\end{align}
Here, $f$ is the scale of self-interaction, known as the axion decay constant. The overdot represents time differentiation, $\nabla^2$ is the usual flat-space Laplacian, and $V'$ represents differentiation of the potential $V(\phi/f)$ with respect to the field $\phi$. An oscillon is a finite-energy solution of \eqref{eq:EOM} that is quasibound by self-interactions. In 3+1 dimensions, which is the focus of our study, all known oscillons have a finite lifetime because they radiate classical scalar waves. To understand whether a potential hosts cosmologically relevant oscillons, one needs a robust computational formalism for obtaining these classical radiation rates. Here, we introduce the physical quasibreather formalism for computing the oscillon radiation and lifetime, while leaving the more technical details to appendix \ref{sec:ThePhysicalQuasibreatherFormalism}.

A physical potential $V$ represents interactions between an integer number of particles, and therefore possesses a well defined Taylor series. Consequently, a field oscillating at fundamental frequency $\omega$ will only couple to integer multiples of $\omega$. Thus, one may look for quasibreather solutions: spherically symmetric, exactly periodic solutions of the equation of motion \eqref{eq:EOM} of the form
\begin{align}
\label{eq:QB1}
    \theta_\te{QB}(t,r,\vartheta,\varphi)\equiv\f{\phi}{f} = \sum_{n\in\N_0}S_n( r,\omega)\sin (n\omega t + \delta_n)\,,
\end{align}
where $\delta_n$ are constant phases, with $\delta_1 = 0$ by the choice of a time coordinate. The harmonic profiles $S_n(r,\omega)$ divide into bound modes $n < m/\omega$ and radiative modes $n > m/\omega$. Solutions of this form were first introduced in \cite{piette1997metastable} and have since been used throughout the oscillon literature to obtain approximate oscillon solutions (see \cite{fodor2019review} for a complete review). Although \eqref{eq:QB1} is a periodic solution of the equations of motion, it is not an infinitely long-lived oscillon; the far-field tails of the radiative harmonics $S_{n > m/\omega}$ decay like $r^{-1}$, and therefore contribute infinite energy.

These unphysical, infinite energy radiative tails have been problematic when interpreting quasibreathers as approximate oscillons. Furthermore, finding a quasibreather of a specific frequency is underdetermined: there are as many different quasibreathers of frequency $\omega$ as there are radiative degrees of freedom, representing the choice of central amplitudes $S_{n>m/\omega}(r = 0,\omega)$. One proposal to resolve this ambiguity is to pick the quasibreather with the minimum radiation amplitude, in an attempt to minimize the influence of the unphysical radiation tails (see e.g. \cite{fodor2019review}). Here, we introduce a different criterion for choosing the quasibreather closest to a physical oscillon. Instead of demanding that the radiative tails are minimized, we will require that the quasibreather is perturbatively close to a radiating solution of \eqref{eq:EOM}.

To this end, we introduce the \emph{orthogonal deformation} (OD)
\begin{align}
\label{eq:OD}
    \theta_\te{OD}(t,r,\vartheta,\varphi)\equiv \sum_{n\omega>m}c_n( r,\omega)\cos (n\omega t + \delta_n)\,,
\end{align}
whose temporal phase is 90 degrees offset from that of the quasibreather \eqref{eq:QB1}. Note that the sum over $n$ only includes the frequencies corresponding to modes with radiative tails, $n\omega > m$.
When added to the standing wave quasibreather \eqref{eq:QB1}, the orthogonal deformation allows for travelling modes (see figure \ref{fig:DPQBAnatomy}). We then define the family of \emph{physical quasibreathers} (PQB), $\theta_\te{PQB}$, parametrized by $\omega$, as those quasibreathers which may be orthogonally deformed $\theta_\te{PQB}\to\theta_\te{PQB} + \theta_\te{OD}$ to satisfy purely outgoing boundary conditions at leading order in $\theta_\te{OD}$ (i.e. $\theta_\te{PQB} + \theta_\te{OD}$ must satisfy the Sommerfeld radiation condition \cite{schot1992eighty}). Note, we will use subscripts to refer either to a general quasibreather $\theta_\te{QB}$ or to a physical quasibreather $\theta_\te{PQB}$ with an OD partner that together satisfy the Sommerfeld radiation condition.

The radiative boundary conditions are enforced at spatial infinity, where the wave equation \eqref{eq:EOM} is well approximated by the Klein-Gordon equation. In this region, the OD and the radiative tails of the PQB are of the same amplitude because they represent purely outgoing radiation. Because $\theta_\te{PQB}$ is a solution of the equations of motion, the perturbation $\theta_\te{OD}$ must backreact at second order ${\mathcal O}(\theta_\te{OD}^2)$, \emph{and} it must obey a homogeneous linear equation to the same order \footnote{This is technically only true once one introduces $\delta\theta$ in appendix \ref{sec:ThePhysicalQuasibreatherFormalism}.}. Therefore, a PQB with small radiative tails must have an OD that is small \emph{everywhere}, compared to the PQB central amplitude. The infinite lifetime limit is the limit of no radiation, and in this case, the PQB approaches a finite energy oscillon. Therefore, the PQB will be the central object in our study of long-lived oscillons.

To summarize, the following three objects are pointwise close to one another: the finite energy oscillon, the PQB, and the orthogonally deformed PQB. This proximity forms the basis of an expansion of the oscillon, which we fully develop in appendix \ref{sec:ThePhysicalQuasibreatherFormalism}, where the oscillon is understood to be a stable perturbation of the orthogonally deformed PQB. As such, oscillon properties (instantaneous frequency, stability, radiation, etc.) may be understood as originating from the nearest PQB. Furthermore, we develop the Floquet analysis of linear perturbations to the deformed PQB in appendix \ref{sec:FloquetAnalysis}. Although linear stability turns out to be a sufficient criterion for the existence of an oscillon, it is not a necessary condition, since stable orbits can (and do) emerge at higher orders. In other words, linear instability does \emph{not} imply the dissolution of the oscillon, since nonlinearities control the size of the linearly unstable perturbations. This effect has important phenomenological consequences for the nature of the oscillon evolution (for examples, see figures \ref{fig:SG},\ref{fig:phi4Power},\ref{fig:SGLinStab},\ref{fig:SGEvo}). Specifically, slow quasiperiodic oscillations around the PQB profile emerge in linearly unstable regions, with amplitude that depends strongly on initial conditions.

Below, section \ref{sec:simpleModes} provides a minimal technical review of our framework, which will be useful in understanding the qualitative features of oscillon evolution in section \ref{sec:lifecycle}. Afterwards, section \ref{sec:numericalWorkflow} outlines the steps in the numerical workflow of computing the PQB and OD, as well as the associated oscillon properties such as lifetime.

\subsection{\label{sec:simpleModes}The mode equations}
At each stage of its life-cycle, the oscillon may be viewed as close to a particular physical quasibreather. This description becomes increasingly precise in the infinite lifetime limit, where radiation goes to zero and the oscillon evolves slowly. Because the oscillon spends a long time in the vicinity of a particular physical quasibreather, the notion of the instantaneous frequency $\omega$ becomes well defined.
Physically, $\omega$ then behaves like an adiabatic parameter, although formally it serves as an index to label which physical quasibreather the oscillon is closest to at a given time. The fact that the oscillon does remain close to the physical quasibreather family is a consequence of its attractive properties, which we make precise in appendix \ref{sec:FloquetAnalysis}.

We are now in position to introduce the mode equations, which describe the spatial profile of the physical quasibreather at a given frequency $\omega$. In the interest of a pedagogical introduction, we will consider the particularly simple case of a single bound harmonic $S_1$ for a potential with parity $V(\theta)= V(-\theta)$, and we will keep only the first radiative harmonic $S_3$.

As outlined above, the potential $V$ is Taylor expandable, and therefore factorizes into a sequence of integer harmonics of the fundamental frequency $\omega$. By restricting to $V(\theta) = V(-\theta)$, only the odd harmonics are coupled to one another, allowing for the following decomposition
\begin{align}
    V'(\theta_\te{PQB})&\equiv m^2 f\sum_{n = 1,3} V_n'(S_1,S_3) \sin (n\omega t) + \dots\,,\\
    V''(\theta_\te{PQB})&\cos(n'\omega t)\\\notag&\equiv m^2\sum_{n = 1,3} V_{n,n'}''(S_1,S_3) \cos (n\omega t) + \dots\,,
\end{align}
where the dots refer to terms proportional to higher frequencies $n\omega$, and terms that contain the small harmonics $S_n$, $n\geq 5$. 
Inserting the quasibreather and the orthogonal deformation into the equations of motion, we arrive at the orthogonally deformed mode equations
\begin{align}
    \label{eq:ModeEQ1}
    \notag
    0&= S_1'' + \f{2}{r}S_1' +\omega^2S_1- m^2 V_{1}'(S_1,S_3)\,,\\
    0&= S_3'' + \f{2}{r}S_3' + (3\omega)^2S_3 - m^2 V_{3}'(S_1,S_3)\,,\\\notag
    0&= c_3'' + \f{2}{r}c_3' + \p{(3\omega)^2-m^2 V_{3,3}''(S_1,S_3)}c_3\,.
\end{align}
To fully specify the solution to this system, we must provide 6 boundary conditions: regularity at the origin
\begin{align}
\label{eq:regularitySimple}
    0&=S_1'(0) = S_3'(0) = c_3'(0)\,,
\end{align}
regularity at spatial infinity
\begin{align}
\label{eq:regularityS1Simple}
    0&=S_1(\infty)\,,
\end{align}
and radiative boundary conditions \cite{schot1992eighty}
\begin{equation}
\begin{aligned}
    \label{eq:RBC}
    0&=\lim_{r\to\infty}\partial_r r S_3(r) - \sqrt{(3\omega)^2 - 1}  r c_3(r)\,, \\
    0&=\lim_{r\to\infty}\sqrt{(3\omega)^2 - 1} r S_3(r) +  \partial_r r c_3(r) \,.
\end{aligned}
\end{equation}
To understand these equations, it is helpful to visualize the evolution of $S_1$ and $S_3$ as the coordinates of a point particle rolling down a hill, where $r$ is now the time coordinate, and the initial stationary particle is placed so that it arrives at the saddle located at the origin when $r\to\infty$ (see figure \ref{fig:rolling}). Out of the continuum of quasibreather initial conditions $S_1(0),S_3(0)$ satisfying this constraint, the orthogonal deformation selects only one, corresponding to the PQB.

\subsection{\label{sec:numericalWorkflow}Calculation workflow}

Here we review the workflow of estimating the oscillon lifetime in the physical quasibreather framework, leaving a more detailed presentation to the appendices.
\begin{enumerate}
    \item The harmonics $S_n$ of the PQB may be thought of as existing in two categories. The perturbative harmonics are those $S_n$ whose amplitude is everywhere small enough that self-interaction can be safely neglected. Those $S_n$ for which this is not true are called non-perturbative. Typically, only a few non-perturbative harmonics are needed to achieve numerical convergence. The physical intuition for whether a harmonic may be treated perturbatively or not is whether it contributes significantly to the binding energy compared to the flux radiated per cycle. In other words, a good rule of thumb for whether a harmonic is perturbative is whether its central amplitude is significantly larger than the leading orthogonal deformation at the origin.
    \item The non-perturbative harmonics (which must include $S_1$) are calculated using a shooting technique, in which the $S_n$'s are propagated from the origin to an outer boundary at $r = r_\te{out}$. At this point, the Sommerfeld radiation condition \eqref{eq:RBC} is used to calculate the OD, $c_n(r_\te{out})$ and $c_n'(r_\te{out})$. From these final conditions, the $c_n$ are propagated back to the origin in the background of the non-perturbative $S_n$. One then checks whether the backwards propagated $c_n$'s satisfy regularity at the origin. We perform a search over initial conditions $S_n(0)$ until regularity is satisfied for all $c_n$'s.
    \item Having computed the non-perturbative harmonics, an arbitrary number of perturbative harmonics may be computed to linear order by solving a sparse matrix equation. In other words, once the hard work of computing the non-perturbative harmonics is done, one may compute the full spectrum of the oscillon to arbitrary harmonic order with little computational cost. One may then re-shoot the non-perturbative harmonics in the background of the perturbative harmonics to account for linear back-reaction, repeating until converged.
    \item The result of these calculations is a semi-non-perturbative expression for the physical quasibreather $S_n$ and its orthogonal deformation $c_n$. The radiation power in each harmonic is easily computed as $P_n = 2\pi r^2 (n\omega)\sqrt{(n\omega)^2 - 1}\p{S_n^2 + c_n^2}$ evaluated at the outer boundary. The sum $\sum_n P_n$ is the PQB approximation to the total power $P$ radiated by the oscillon.
    \item Having calculated the outgoing power $P$ as a function of the PQB frequency $\omega$, we may approximate the lifetime of the oscillon near the physical quasibreather trajectory as $T = \int\diff\omega (dE_B/d\omega)/P$, where $E_B$ is the \emph{bound} energy in the oscillon, defined as the difference between the PQB and OD energy (see appendix \ref{sec:EnergeticInstability}).
\end{enumerate}
We provide a public implementation of this protocol for the case of a single non-perturbative harmonic in potentials with parity --- a fast and easy-to-use tool to obtain ballpark estimates of oscillon properties at larger frequencies \footnote{\faGithub\href{https://github.com/SimpleOscillon/Code}{\hspace{0.1cm}Simple oscillon code.}}.

\section{\label{sec:lifecycle}The oscillon life-cycle}

Here we review and expand upon previous literature results \cite{fodor2006oscillons,saffin2007oscillons,fodor2008small,gleiser2009general,fodor2009radiation,amin2010flat,amin2012oscillons,salmi2012radiation,andersen2012four,mukaida2017longevity,visinelli2018dilute,fodor2019review,gleiser2019resonant,olle2020oscillons,zhang2020classical,olle2020recipes} in order to identify the main mechanisms responsible for oscillon longevity and death. We point out two distinct effects contributing to oscillon longevity: \emph{geometric decoupling} and \emph{destructive interference}, both of which may be thought of together as the \emph{form-factor} of the oscillon coupling to radiation. It is important to separate form-factor into these two effects because they intervene at different times, and have different consequences for oscillon evolution. Often, an oscillon's lifetime is dominated by one mechanism or the other, while the longest lived oscillons take advantage of both simultaneously. Separately, as the oscillon ages and grows more diffuse, it will inevitably undergo an \emph{energetic death}, beyond which its energy would be forced to unphysically increase. These three effects are all pointed out in figure \ref{fig:genericPower}, which depicts the typical radiation history of an oscillon. Below, we provide a semi-quantitative overview of these three effects.

\subsection{\label{sec:geometricDeco}Geometric decoupling}

Recall that the oscillon is a smooth, nearly coherent object, coupling to integer multiples $n$ of its fundamental frequency $\omega$ through many-to-one interactions at leading order  $\phi^{n + 1}$. As the oscillon radiates binding energy throughout its life, its fundamental frequency increases towards $m$ (see figure \ref{fig:timeCurve}), and its typical size $2\pi/\sqrt{m^2 - \omega^2}$ blows up, where $\sqrt{m^2-\omega^2}$ is the binding energy per particle. Therefore, a natural separation of scales occurs between the length scale of radiation $2\pi/n\omega$ and the size of the oscillon, leading to an exponential suppression of the oscillon's coupling to radiative modes $n\omega$, $n\geq 2$. According to a standard Riemann-Lebesgue suppression argument, the ratio of the $n\omega$ harmonic amplitude to the fundamental harmonic central value scales as $\gamma^n$, with
\begin{align} 
     \gamma\approx \exp\ps{-G\f{\omega}{\sqrt{m^2 - \omega^2}}}\,,
\end{align}
where $G$ is an order 1 geometrical factor, used here as a stand-in for the exact shape of the oscillon. The fact that the geometrical factor $G$ is in the exponent shows that even modest changes in the oscillon's shape can dramatically change its lifespan, emphasizing the importance of accurately resolving the oscillon geometry. Moreover, because the factor $\omega/\sqrt{m^2 - \omega^2}$ becomes larger as $\omega$ approaches $m$, the differences between potentials will be exaggerated in this limit, while low-frequency oscillons will typically be similar to one another (see figures \ref{fig:SG} and \ref{fig:monodromy} for an example).
As a consequence of this growing separation of scales, oscillons whose frequency $\omega$ approaches the mass $m$ radiate at increasingly suppressed rates, so that the last phase of the oscillon's life is often the longest. We refer to this general trend as geometric decoupling.

\subsection{\label{sec:simpleConfinement}Destructive interference and the minimum radiation condition}

\begin{figure}
    \centering
    \includegraphics[width = \columnwidth]{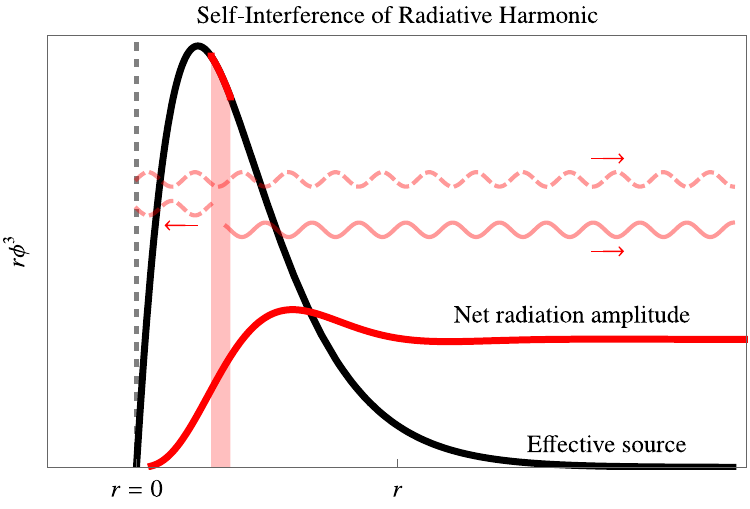}
    \caption{
    A physical model for an oscillon radiating into the third harmonic. The black line represents the background oscillon source $\phi^3$, while the red lines represent the amplitude of the radiated field. The spherical symmetry of the oscillon imposes boundary conditions at the origin which behave like an optical mirror: an inward propagating spherical wave is reflected at the origin, propagating back outward with the opposite phase. The result is that the oscillon radiation may experience two kinds of self-interference: interference from the physical extent of the source, analogous to diffraction of a laser beam through a finite-width slit, and interference due to the spherical symmetry of the oscillon, represented by the mirror. At certain oscillon frequencies, these two effects conspire to destructively interfere, trapping a nominally free harmonic.}
    \label{fig:Diffraction}
\end{figure}

Throughout the oscillon lifetime, radiative harmonics are subject to self-interference, which is totally destructive at exceptional frequencies. At these points, destructive interference completely confines specific harmonics, and subverts the expected radiation hierarchy implied by geometric decoupling. When the leading harmonic is confined, the overall radiation amplitude shrinks by another global factor of $\gamma$. For many especially long-lived oscillons, a period near harmonic confinement dominates the total lifetime. In principle, it is possible to imagine engineering ultra-long-lived oscillons by aligning the destructive interference of multiple harmonics, leading to additional suppression by $\gamma^\ell$, where $\ell$ is the number of aligned exceptional points. In practice, these constructions are necessarily fine tuned, since each resonance must be aligned to order $\gamma^{\ell - 1}$.

\subsubsection{Interferometric analogue}
The basic physics of oscillon radiation is captured by the physical model in figure \ref{fig:Diffraction}, which describes an interference experiment reminiscent of the classic Lloyd's mirror. In this simple one-dimensional setup, a coherent, finite-sized, optical source at $r>0$, representing the oscillon's coupling to the radiative harmonic, is placed in front of a mirror at $r=0$, representing the spherical symmetry of the oscillon. Each point in the source experiences interference both from its reflection, and from its neighbors. Let the spatial location and magnitude of the source be described by $\tilde {\mathcal J}(r)$. The direct radiation reaching the observer is therefore
\begin{equation}
    A_\text{direct}(t, r) = \int_0^\infty \diff x\,\tilde {\mathcal J}(x)\,e^{i[\omega t - k(r-x)]}\,.
\end{equation}
On the other hand, the reflected light paths sum up to an amplitude:
\begin{equation}
    A_\text{reflected}(t,r) = \int_0^\infty \diff x\,\tilde {\mathcal J}(x)\,e^{i[\omega t - k(r+x) + \pi]}\,,
\end{equation}
where, crucially, a half-wavelength path difference is picked up upon reflection at the mirror. This is equivalent to enforcing the usual regularity conditions at the origin in a spherically symmetric field solution. Finally, the observer adds up these contributions coherently, which explicitly leads to an amplitude equal to the sine-transform of the source:
\begin{equation}
\begin{aligned}
    A_\text{obs}(t,r) &= A_\text{direct}(t,r) + A_\text{reflected}(t, r)\,, \\
    &= 2e^{i[\omega t - kr + \pi/2]}\int_0^\infty \diff x\,\tilde {\mathcal J}(x)\,\sin(kx)\,.
\end{aligned}
\end{equation}
In the following section, we derive a similar result from the mode equations of the PQB, and quantify corrections to this simplified picture.

\subsubsection{\label{sec:DetailedConfinement}The physical quasibreather picture}

In the previous section, we introduced a simple interpretation of the oscillon radiation in terms of the interference of a coherent source with its own reflection. Here we study the mode equations \eqref{eq:ModeEQ1}, in which the first radiative harmonic $S_3$ and the orthogonal deformation $c_3$ are treated as a perturbation of the fundamental $S_1$. Under this perturbative assumption, the mode equations for the radiative harmonic $\tilde S_3 \equiv rS_3$ and its orthogonal deformation $\tilde c_3\equiv r c_3$ further simplify to the frictionless linear system
\begin{align}
    \tilde S_3''(r) +  k_S^2(r)\tilde S_3(r) &= r{\mathcal J}_3(r)\,, \\
    \tilde c_3''(r) +  k_c^2(r)\tilde c_3(r) &= 0\,.
\end{align}
Here, $k_S$ and $k_c$ represent the $r$-dependent wavenumbers of the third harmonic $S_3$ and $c_3$ in the background of the fundamental harmonic $S_1$, and $r{\mathcal J}_3(r)$ corresponds to the $3\to 1$ processes generating the radiation. 
Note, the wavenumber is different for the third harmonic $S_3$ and its orthogonal deformation $c_3$, a distinction explicitly derived in the appendix result \eqref{eq:S3Matrix}. There, we find that the difference between $k_S$ and $k_c$ appears at sixth order in a Bessel expansion of the background, and therefore is typically small, making the approximation $k_S = k_c$ quantitatively good in most circumstances. 
We can solve the linear system analytically in terms of two linearly independent solutions $y_{1,2}^S(r)$ and $y_{1,2}^c(r)$ of the {\em homogeneous} equations. In this case, the expression for the Green's function is simple and the full solution becomes a sum of homogeneous (defined by initial conditions) and inhomogeneous contributions, of the form:
\begin{align}
\notag
    \label{eq:HomogeneousInhomogeneousSk}
    \tilde S_3(r) =& a_1^\te{H}y_1^S(r) + a_2^\te{H}y_2^S(r) + \\
    &+ y_1^S(r)\int_0^r \diff r'\,r'\,{\mathcal J}_3(r')\,\frac{y_2^S(r')}{W^S(r')} \\\notag
    &- y_2^S(r) \int_0^r \diff r'\,r'\,{\mathcal J}_3(r')\frac{y_1^S(r')}{W^S(r')}\,, \\
    \tilde c_3(r) =& b_1^\te{H}y_1^c(r) + b_2^\te{H}y_2^c(r)\,,
\end{align}
where $W^S(r)\equiv y_1^S(r){y_2^S}'(r) - {y_1^S}'(r)y_2^S(r)$ is the Wronskian. Let $y_1^{S,c}$ be the sine-like solution (nonzero derivative at $r=0$) and let $y_2^{S,c}$ be the cosine-like solution (zero derivative at $r=0$). Regularity at the origin requires that only sine-like initial conditions are allowed, constraining the cosine-like terms to be zero $b_2^\te{H} = a_2^\te{H} = 0$.

In the far-field region, all solutions $y_{1,2}^{S,c}$ are simple combinations of sines and cosines of frequency $k_3=\sqrt{(3\omega)^2-m^2}$. However, orthogonality between $y_1$ and $y_2$ is generally not maintained into the far-field. Without loss of generality, we can introduce phase-shifts to express these misalignments,
\begin{align}
    y_1^S &= \sin (k_3 r)\,,\\
    y_2^S &= \cos (k_3 r + \varphi_2^S)\,,\\
    y_1^c &= \sin (k_3 r + \varphi_1^c)\,,
\end{align}
with the understanding that when these phase-shifts are zero, we regain the simple constant-wavenumber Helmholtz solutions. These phase-shifts can in principle be computed in the WKB approximation. Furthermore, we define the orthogonal components $y_s = \sin(k_3 r)$ and $y_c = \cos(k_3 r)$ against which we can project the shifted solutions, leading to
\begin{align}
    y_1^S &= y_s\,,\\
    y_2^S &= y_c\cos\varphi_2^S - y_s\sin\varphi_2^S\,,\\
    y_1^c &= y_c\sin\varphi_1^c + y_s\cos\varphi_1^c\,.
\end{align}
Substituting, we collect the orthogonal contributions to the radiative tails as
\begin{align}
    \label{eq:HomInhomLimitS}
    \notag
    S_3(r)r &\xrightarrow[r\to\infty]{} y_s(a_1^\te{H} + a_1^\text{I} + a_2^\text{I}\sin\varphi_2^S) - y_c a_2^\text{I}\cos\varphi_2^S\,, \\
    c_3(r)r &\xrightarrow[r\to\infty]{} y_s  b_1^\te{H}\cos\varphi_1^c + y_c b_1^\te{H}\sin\varphi_1^c\,,
\end{align}
where $a_1^\te{I}$ and $a_2^\te{I}$ are fixed, representing the total inhomogeneous contribution from the oscillon background
\begin{align}
    a_1^\te{I}&=\int_0^\infty \diff r'\,r'\,{\mathcal J}_3(r')\,\frac{y_2^S(r')}{W^S(r')} \,, \\
    a_2^\te{I}&=\int_0^\infty \diff r'\,r'\,{\mathcal J}_3(r')\frac{y_1^S(r')}{W^S(r')}\,.
\end{align}
Radiative boundary conditions \eqref{eq:RBC} match the coefficients of $y_s$ and $y_c$ between $S_3$ and $c_3$, which uniquely determines the homogeneous degrees of freedom,
\begin{equation}
\begin{aligned}
    \label{eq:homogeneousRBC}
    b_1^\te{H} &= a_2^\te{I}\cos\varphi_2^S\sec\varphi_1^c \\
    a_1^\te{H} &= -a_1^\te{I} + a_2^\te{I}(-\sin\varphi_2^S + \cos\varphi_2^S\tan\varphi_1^c).
\end{aligned}
\end{equation}
Consequently, the solution simplifies to
\begin{align}
    S_3(r) r &= a_2^\te{I}\cos(\varphi_2^S)(y_s\tan\varphi_1^c - y_c)\,,\\
    c_3(r) r &= a_2^\te{I}\cos(\varphi_2^S)(y_s + y_c\tan\varphi_1^c)\,.
\end{align}
In other words, the amplitude of the radiation is always proportional to the inhomogeneous contribution $a_2^\te{I}$. At exceptional frequencies, this contribution is exactly zero and the harmonic experiences totally destructive interference; this is visible in the power versus frequency plots as a sudden drop (see figures \ref{fig:SG} and \ref{fig:monodromy} for example). Therefore, in this linear model the condition for totally destructive interference is
\begin{align}
    0&=\int_0^\infty \diff r'\,r'\,{\mathcal J}_3(r')\frac{y_1^S(r')}{W^S(r')}\,.
\end{align}
In the case of a flat wave-number, i.e. Helmholtz system, this is precisely the sine-transform of the source, as predicted by the simple interferometric model. 
Because totally destructive interference is equivalent to a single constraint on one free parameter $\omega$, we conclude this effect is \emph{generic}, and not the result of some fine tuning.

To reach this result, we have effectively solved for the physical quasibreather, defined by the choice of $a_1^\te{H}$ in \eqref{eq:homogeneousRBC}, at the level of the third harmonic and in a linear approximation. In previous literature (e.g. \cite{fodor2019review}), a different quasibreather was highlighted as relevant in approximating the oscillon, namely the {\em minimum-radiation quasibreather}. This corresponds to a different choice of homogeneous parameters; in this case, the construction of $c_3$ is irrelevant and the value of $a_1^\te{H}$ is chosen such that $S_3$ is minimized at the level of \eqref{eq:HomInhomLimitS}, specifically by picking:
\begin{equation}
    a_1^\te{H} = -a_1^I - a_2^I\sin\varphi_2^S\,.
\end{equation}
We see that this differs from the physical quasibreather answer \eqref{eq:homogeneousRBC} by an additional $a_2^\te{I}\cos\varphi_2^S\tan\varphi_1^c$, which is zero in the case when $\varphi_1^c=0$, i.e. when the wavenumbers $k_S(r)$ and $k_c(r)$ are identical functions of $r$. While typically small, differences between $k_S(r)$ and $k_c(r)$ appear at higher-orders in the background, and are not guaranteed to be perturbative --- as derived below in appendix \ref{sec:PerturbativeHarmonicFormulae}. Therefore, the minimum-radiation quasibreather and the physical quasibreather are generally close but distinct, and are identical only at the exceptional `dip' frequency where both predict zero radiative tails.

\subsection{\label{sec:EnergeticDeath}Energetic death}
As explained in section \ref{sec:geometricDeco}, the spatial extent of the oscillon increases as it radiates away its binding energy. On the other hand, the balance between self- and binding-energy demands that the oscillon's central amplitude decreases. Depending on the number of spatial dimensions, one effect or the other dominates the oscillon's total energy as $\omega$ approaches $m$. In particular, in three or more spatial dimensions, the volume turns out to grow faster than the central amplitude shrinks. The oscillon's parent PQB also obeys the same scaling relation, and at some point the bound energy in the PQB will necessarily begin to increase. To keep up, the oscillon would need a source of energy; in its absence, the oscillon is forced off the PQB trajectory, in a process we call energetic death.

To make these ideas precise, we can invoke the mode equations \eqref{eq:ModeEQ1}, in the limit of small central amplitude $S_1(0)$. Note, because the oscillon's volume is large, it is geometrically decoupled from radiation according to the argument in section \ref{sec:geometricDeco}, and therefore it is safe to neglect backreaction from the radiative harmonics. Keeping only the leading quartic nonlinearity in the potential, $S_1$ is described by
\begin{align}
    0&=S_1'' + \f{d-1}{r}S_1' - (m^2-\omega^2)S_1 + \f{3}{4} m^2\lambda S_1^3\,.
\end{align}
Here, $d$ is the number of spatial dimensions. To extract the scaling of $S_1(0)$, we match the binding energy of the oscillon to its self-energy, leading to
\begin{align}
    (m^2-\omega^2)S_1^2 \sim m^2\lambda  S_1^4\,.
\end{align}
Therefore, the scaling of the central amplitude is independent of dimensions, namely
\begin{align}
\label{eq:amplitudeScaling}
    S_1(0)\propto\sqrt{m^2-\omega^2}\,.
\end{align}
On the other hand, since the spatial extent of the oscillon scales like scale $1/\sqrt{m^2 - \omega^2}$ (as seen in section \ref{sec:geometricDeco}), its volume must increase according to
\begin{align}
    V\sim\p{m^2 - \omega^2}^{-d/2}\,.
\end{align}
Combining these two scalings results in the oscillon's total energy
\begin{align}
\label{eq:energeticDeath}
E\propto V S_1(0)^2 \propto \p{m^2 - \omega^2}^{1-d/2}\,,
\end{align}
which grows as $\omega$ approaches $m$ for spatial dimension $d\geq 3$. In other words, the expectation that the oscillon energy decreases as a function of $\omega$ is only true up to a specific frequency strictly less than $m$. Beyond this point, the oscillon energy is forced to increase as a result of weak self-interaction. Such an increase is unphysical, and the value of $\omega$ at which the PQB's energy is minimized sets the moment of death. For an earlier argument along these lines, see \cite{fodor2009radiation,fodor2019review}.

For an explicit comparison, take the $d = 1$ sine-Gordon oscillon, which has a simple analytic form
\begin{align}
    \phi = 4\arctan\ps{\f{\sqrt{m^2 - \omega^2}\cos\omega t}{\omega\cosh\sqrt{m^2 - \omega^2}x}}\,.
\end{align}
In the $\omega\to m$ limit, the energy of the sine-Gordon oscillon is exactly $16 \sqrt{m^2 - \omega^2}$, which matches our predicted scaling.

All the examples of oscillons studied in sections \ref{sec:prescription}, \ref{sec:tuned?} and \ref{sec:illustrativeExamples} live in three spatial dimensions, and therefore exhibit an energetic death. In other words, for each oscillon there is a specific frequency strictly less than $m$ beyond which the scalar field may no longer exist close to a PQB. After this point, our formalism no longer applies, and the oscillon is considered ``dead.'' Afterwards, gravity may take over leading to the formation of much more diffuse configurations such as axion stars \cite{braaten2018axion,visinelli2018dilute}. In our numerical simulations, this moment of death is distinctly visible as a ``loop,'' representing the rapid conversion of the oscillon into radiation through $3\to1$ processes (see figures \ref{fig:SG}, \ref{fig:phi4Power}, and \ref{fig:QCD}).

\section{\label{sec:prescription}A prescription for oscillon longevity}

Here we provide a procedure for generating potentials that support cosmologically long-lived oscillons. In section \ref{sec:lifecycle}, we explained how the longest-lived oscillons exhibit a combination of geometric decoupling and destructive interference. Geometric decoupling refers to the suppression of radiation when the oscillon size is much larger than the radiation wavelengths, which is especially pronounced at large frequencies $\omega$ close to $m$. For a large oscillon,  the interferometric `fringe pattern' also occurs more rapidly, leading to more instances of destructive interference which further suppresses radiation. Thus, we may find long-lived oscillons by searching for potentials that support large oscillons at frequencies $\omega$ close to $m$. An apparent obstacle to this goal is due to energetic death (see section \ref{sec:EnergeticDeath}), which limits the frequencies for which the oscillon can have decreasing energy as a function of $\omega$. In the following, we identify a feature in the scalar potential that can stave off energetic death \emph{and} produce large oscillons.

In section \ref{sec:physicalQuasibreatherMinTech}, we introduced the mode equations \eqref{eq:ModeEQ1} obeyed by the radial profiles of the PQB harmonics $S_n(r)$, and the sense in which these harmonics may be thought of as the coordinates of a point particle, whose initial condition is tuned so that $(S_1,S_3,\dots) = \vec 0$ at $r = \infty$. Here we aim to study the longest-lived oscillons, whose radiation is necessarily small. Moreover, we will focus on large frequencies $\omega\approx m$, for which higher harmonics are further suppressed by a natural separation of scales $\omega\gg\sqrt{m^2 - \omega^2}$. Therefore, we will drop the higher harmonics $n\geq 3$ in this section's analysis, and work with a simplified 1-dimensional point-particle picture, representing the radial profile of the fundamental mode, $S_1(r)$.

We now introduce the equations that govern $S_1$ from first principles, using an effective action technique, equivalent to the PQB formalism for a single bound harmonic. The Lagrangian describing the real scalar $\phi$ is
\begin{align}
    L[\phi] &= \int\diff^3 x \ps{\f12 \dot\phi^2 - \f12\nabla^2\phi^2 - V(\phi)}\,.
\end{align}
Because both $V$ and $\phi$ are proportional to $f^2$ (as in \eqref{eq:PeriodicSum}), $f$ is an overall factor in the action, and therefore does not contribute to the dynamics. Hence, for the rest of this section, we will work in units of $f = 1$.
Since we are looking for quasiperiodic, spherically symmetric solutions dominated by the fundamental mode, we substitute $\phi= S_1(r)\sin\omega t$ and integrate out time, leading to the effective action for $S_1$:
\begin{align}
    S_{\te{eff}}[S_1]&=\int_0^{\f{2\pi}{\omega}}\diff t L[S_1\sin\omega t]\,,\\
    &=-\f{\pi}{\omega}\int 4\pi r^2\diff r\ps{\f12{S_1'(r)}^2 -  V_\te{eff}(S_1)}\,.
\end{align}
By integrating out time, we arrive at an action for a point particle $S_1(r)$, where $r$ acts like a time coordinate, and the resulting effective potential is
\begin{align}
    V_\te{eff}[S_1]&\equiv \f12\omega^2 {S_1(r)}^2 - \f{\omega}{\pi}\int_0^{\f{2\pi}{\omega}}\diff t\, V(S_1 \sin\omega t)\,.
\end{align}
Finally, the equation of motion for $S_1$ arising from this effective action carries a $2/r$ friction term from the spherical Jacobian
\begin{align}
\label{eq:effectiveS1}
    0&= S_1'' + \f{2}{r}S_1' + V_\te{eff}'(S_1)\,,
\end{align}
where $V_\te{eff}'$ represents the derivative of $V_\te{eff}$ with respect to $S_1$.

A solution of these equations which describes an oscillon profile needs to respect regularity conditions at $r = 0$ and $r = \infty$, corresponding to $S'(0) = 0$ and $S(\infty) = 0$. All solutions which respect regularity at $r = \infty$ must exponentially decay, since $V_\te{eff}$ behaves like a quadratic hilltop $-\f12(m^2 - \omega^2) S_1^2$ for small $S_1$. From the perspective of the point particle, this means that initial conditions are tuned such that $S_1$ has just enough energy to climb up the hilltop at 0.

In order to engineer large oscillons, we need the particle $S_1$ to stay at small velocities so that the oscillon interior spreads out. Initializing on a hillside of $V_\te{eff}$ is detrimental to this goal, since the slope of $V_\te{eff}$ controls the speed of $S_1$, typically leading to a small oscillon core. On the other hand, releasing $S_1$ close to a hilltop allows $S_1$ to remain at low velocities for a time inversely proportional to the initial displacement of $S_1$ from the hilltop.

\begin{figure}
    \centering
    \includegraphics[width = \columnwidth]{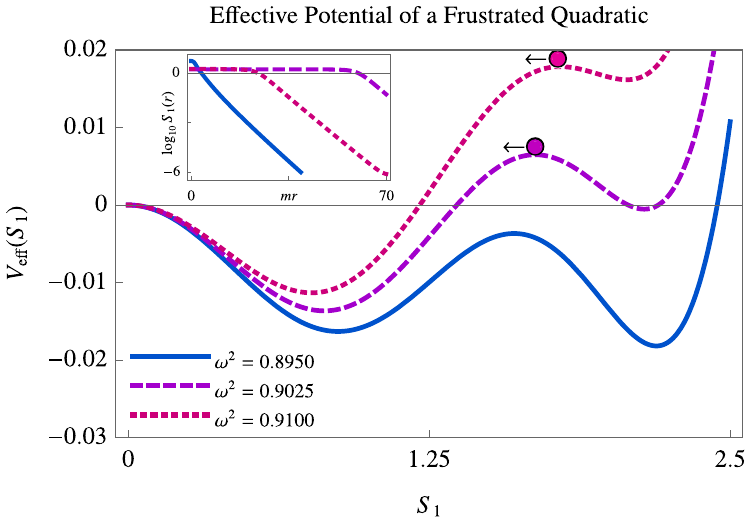}
    \caption{Effective potential $V_\te{eff}(S_1)$ for a long-lived oscillon, at three nearby frequencies. The example is obtained using the frustrated quadratic method defined in \eqref{eq:frustratedObjective} with $m_f^2 = 0.9 m^2$ and $b = 2$, computed using three Fourier coefficients ${\mathcal V}_{1,2,3}$ with ${\mathcal V}_3$ forced to satisfy the mass constraint in \eqref{eq:PeriodicSum}. We see that as $\omega$ passes through the frustrated mass $m_f$, new solutions to the equations of motion \eqref{eq:effectiveS1} emerge, specifically when the local maximum of the effective potential increases to positive values. The balls are placed at the values $S_1(0)$ which initialize physical oscillon solutions at the respective frequencies $\omega$. The inset figure shows the trajectories of the smallest-amplitude solutions of \eqref{eq:effectiveS1} for each of the three potentials plotted.}
    \label{fig:frustrationSchematic}
\end{figure}

\begin{figure}
    \centering
    \includegraphics[width = \columnwidth]{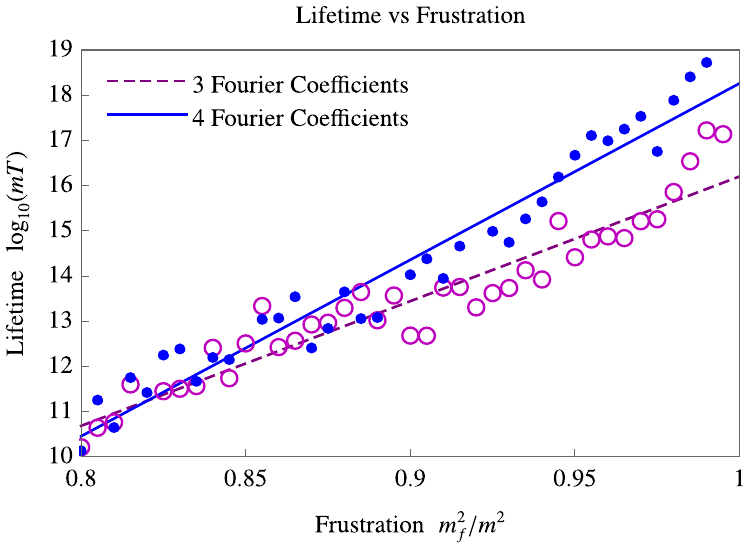}
    \caption{Lifetime versus frustration for oscillons in frustrated quadratic potentials, computed using three and four Fourier coefficients (see equation \eqref{eq:frustratedObjective}). The lifetimes are integrated over the interval $\omega\in[0.8,0.999]$ in the one-non-perturbative harmonic PQB formalism. We speculate that introducing more Fourier coefficients leads to longer-lived oscillons, since the frustration mass can be closer to $m$ before self-interactions become repulsive, leading to enhanced geometric decoupling. The line of best fit for three coefficients (dashed purple) is $\log_{10}(m t) = 28(m_f/m)^2 -11$, and the best fit with four coefficients (solid blue) is $\log_{10}(m T) = 39(m_f/m)^2 -21$.}
    \label{fig:frustrationLifetime}
\end{figure}
Therefore, we need to connect hilltop-initialized solutions (i.e. low central velocity) to physical solutions (i.e. which arrive at $S_1(\infty) = 0$). To compensate for the energy lost by $2/r$ friction, a physical solution must be initialized with positive potential energy (where we've normalized $V_\te{eff}(0) = 0$). Effective potentials which satisfy these two conditions will have non-trivial local maxima, whose hilltop is higher than 0 (see figure \ref{fig:frustrationSchematic}). 

Here we reverse engineer a class of scalar potentials $V(\phi)$ which generate effective potentials $V_\te{eff}(S_1)$ with the aforementioned hilltops. This construction makes use of the fact that the mass term in the effective potential $V_\te{eff}(S_1)$ acquires an $\omega^2$ offset compared to the mass term in the scalar potential $V(\phi)$. Based on this observation, we introduce the family of \emph{frustrated quadratics}, whose Fourier coefficients (in the basis expansion \eqref{eq:PeriodicSum}) are chosen as the solution to the following optimization problem:
\begin{equation}
\begin{aligned}
\label{eq:frustratedObjective}
    \te{minimize }\hspace{0.25cm}& \max_{\phi\in[-b,b]} \abs{V(\phi) - \f{1}{2}  m_f^2\phi^2}\,,\\
    \te{subject to }\hspace{0.25cm}&V''(\phi = 0) = m^2\,,
\end{aligned}
\end{equation}
where $0 <b< \pi$, and $0<m_f < m$ is the frustrated mass. In other words, we are forcing the potential to have mass $m$ at small $\phi$, and a different, smaller mass $m_f$ at larger $\phi$. For frequencies $\omega$ close to $m_f$, the effective potential $V_\te{eff}$ will consist of a series of hills and valleys inside the interval $S_1\in[-b,b]$, whose amplitude is controlled by how tightly the objective \eqref{eq:frustratedObjective} is optimized.

As local hilltops in such potentials rise above the zero-potential line, new oscillon solutions emerge. When the hilltop is precisely at the zero-line, this new solution is wholly unphysical, carrying infinite energy. As $\omega$ increases, this hilltop is pushed upwards, and $S_1(0)$ starts with more potential energy that needs to be dissipated through friction. As a consequence, $S_1(0)$ starts further from the hilltop so that it begins rolling earlier while the $2/r$ friction is still active, corresponding an oscillon with a smaller radius and less energy (see the inset of figure \ref{fig:frustrationSchematic}). Even though this branch may appear at very large $\omega$ close to $m$, this effect guarantees there is some finite range of frequencies over which the energy of these solutions decreases, meaning a physical oscillon can be supported.

In figure \ref{fig:frustrationLifetime}, we plot the lifetime of oscillons in frustrated quadratics as a function of the frustration $m_f^2/m^2$. The frustration mass $m_f$ controls the frequency at which new hilltop solutions emerge. As $m_f$ increases towards $m$, the appearance of these new branches occurs at larger frequencies, taking advantage of enhanced geometric decoupling and leading to longer lifetime. Increasing the number of Fourier coefficients in the potential reduces the height of the hilltops in the effective potential, allowing them to emerge at larger frequencies. Further, higher frequencies in the potential pushes the hilltops closer to $S_1 = 0$, allowing for lower-energy oscillons. We speculate that a fixed number of Fourier coefficients in the potential implies an upper bound on oscillon lifetimes, although we leave this question to future work. 

\section{\label{sec:tuned?}Is longevity fine-tuned?}
There are many known examples of potentials which support very long-lived oscillons, including those identified in section \ref{sec:prescription}. However, the precise form of these potentials remains largely unconstrained by a UV theory, and therefore it is not clear how to assess whether their longevity is a result of fine-tuning, since the distribution from which the potential coefficients are sampled strongly influences the lifetime. Therefore, we introduce two notions of tuning that attempt to quantify the difficulty of constructing a theory with a long-lived oscillon:
\begin{enumerate}
    \item \emph{Global Tuning} asks what fraction of parameter space hosts long-lived oscillons. A typical object of study is the probability distribution of lifetimes, sampled with minimal priors over the potential coefficients in some natural basis.
    \item \emph{Local Tuning} asks whether a given long-lived oscillon is sensitive to variation in its potential parameters. The typical objects of study are the local gradient and curvature of the lifetime with respect to the parameter space at the point in question.
\end{enumerate}

In sections \ref{sec:globalTuning} and \ref{sec:localTuning}, we address the genericity of long-lived oscillons in periodic potentials with parity.
The advantage of studying periodic potentials is that they are naturally expanded in the Fourier basis. Without any theoretical priors, a natural scale for the Fourier coefficients is $m^2 f^2$, and variations will be of the same size.

One may also be interested in studying the genericity of long-lived oscillons in monodromy potentials \cite{silverstein2008monodromy,dubovsky2012axion,kaloper2017london}. Since an oscillon has a finite amplitude, one may restrict the aperiodic monodromy potentials to a compact interval, which is fully described by a Fourier expansion. However, any realistic model of axion monodromy is asymptotically a power law, meaning the high frequency modes of the potential are perfectly correlated. To sample the full space of monodromy potentials, one must sample from a distribution that imposes this correlation. In the absence of a reliable way to select coefficients from this distribution, we leave this question to future work. Instead, in section \ref{sec:monodromy}, we scan the one-parameter family of monodromy potentials studied in \cite{amin2012oscillons,zhang2020classical,olle2020recipes}.

\subsection{\label{sec:globalTuning}Global tuning}
\begin{figure}
    \centering
    \includegraphics[width = \columnwidth]{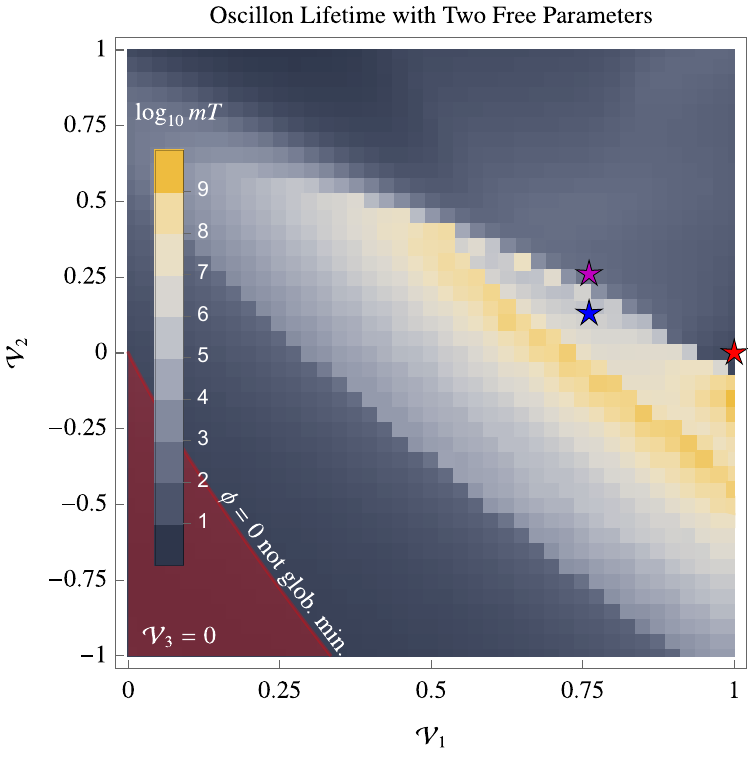}
    \caption{Accessible oscillon lifetimes in a periodic potential with two degrees of freedom ${\mathcal V}_1,{\mathcal V}_2$. Here ${\mathcal V}_3$ is constrained such that the mass is fixed to $m$, with all other ${\mathcal V}_{n\geq4} = 0$. The red region indicates parts of the parameter space where $\phi = 0$ is not a global minimum of the potential, and has significantly shorter lifetimes. The stars indicate potentials for which we have compared our formalism with multiple non-perturbative harmonics to direct numerical simulation (see Figure \ref{fig:SG}). The peninsula of longevity corresponds to the emergence of a frequency at which the third harmonic experiences totally destructive interference at `dips.' The yellow banding corresponds to the migration of dips to higher frequencies, where geometric decoupling suppresses the fifth harmonic, increasing the impact of the dip. At the upper right of these bands, the dip migrates to frequencies higher than that of energetic death, creating a longevity cliff.}\label{fig:globalTuning1}
\end{figure}
\begin{figure}
    \includegraphics[width = \columnwidth]{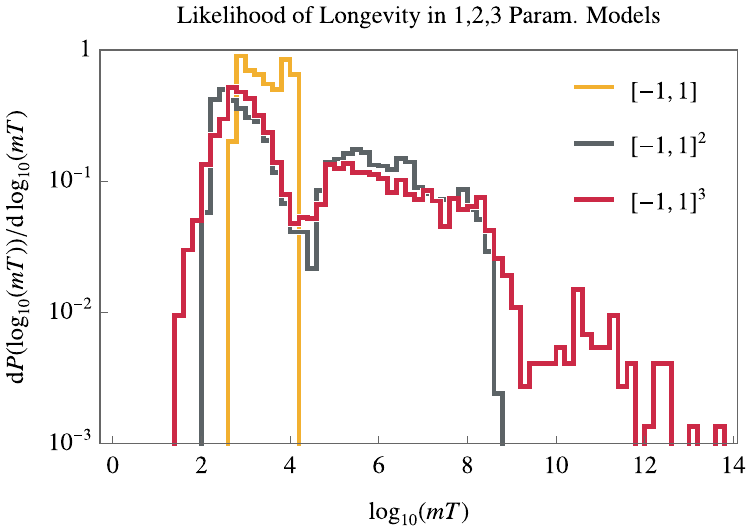}
    \caption{The distribution of oscillon lifetimes for 1 (yellow), 2 (gray), and 3 (red) degrees of freedom in a periodic potential. We uniformly sample the $n_\te{DOF}$-dimensional cube ${\mathcal V}_1,\dots{\mathcal V}_{n_\te{DOF}}\in[-1,1]$ restricting the potential such that $\phi = 0$ is a global minimum, and ${\mathcal V}_{n_\te{DOF} + 1}$ is fixed such that the mass is $m$, with the remaining ${\mathcal V}_n$ set to 0. Lifetimes are computed in the interval $\omega/m\in[0.8,0.995]$ in the single-non-perturbative-harmonic approximation. The geometric suppression of the radiative modes means that these frequencies likely dominate the oscillon lifetime, and that the perturbative radiation approximation is typically good. We see that each new degree of freedom is observed to introduce a new island of longevity (island 1 $\log_{10} m T\in [0,4]$, island 2 $\log_{10} m T\in [4,9]$, island 3 $\log_{10} m T\in [9,14]$).}
    \label{fig:globalTuning2}
\end{figure}

\begin{figure}
    \centering
    \includegraphics[width = \columnwidth]{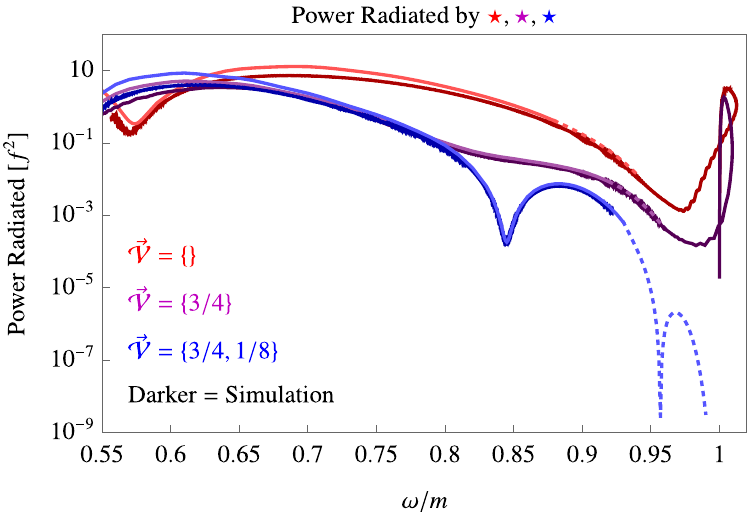}
    \caption{The power radiated by the oscillons in the potentials denoted by stars in figure \ref{fig:globalTuning1}. The dark curves are data from explicit numerical simulations (see appendix \ref{sec:simulation}), while the lighter curves are computed in the PQB formalism. The PQB predictions become dotted in regions of linear instability, as computed using the methods described in appendix \ref{sec:FloquetAnalysis}. Notice that at low frequencies, the oscillon power curves are of similar magnitude, diverging at larger frequencies due to geometric decoupling, as explained in section \ref{sec:geometricDeco}. The loops at the end of the simulations correspond to the oscillon rapidly converting into $3\omega$ modes past the point of energetic death, causing the measured frequency at the origin to briefly grow larger than 1.
    }
    \label{fig:SG}
\end{figure}
In this section, we study the distribution of oscillon lifetimes as a function of the potential coefficients in the Fourier basis. In particular, we will consider periodic potentials with parity
\begin{align}
\label{eq:PeriodicCoeff}
    V(\theta) = m^2 f^2\sum_{n = 1}^\infty \f{{\mathcal V}_n}{n^2}\p{1 - \cos n\theta}\,,\hspace{0.5cm} \sum_{n = 1}^\infty {\mathcal V}_n = 1\,,
\end{align}
where the sum of the coefficients ${\mathcal V}_n$ constrains the mass of $\phi$ to be $m$. In Figure \ref{fig:globalTuning1}, we plot the lifetime versus the free variation of the first two coefficients $n = 1,2$ with the third constrained so that the sum in \eqref{eq:PeriodicCoeff} is satisfied, with all other ${\mathcal V}_n$ set to zero. The mass constraint in \eqref{eq:PeriodicCoeff} naturally sets the typical scale of ${\mathcal V}_n$ to 1. Therefore, we restrict our study to ${\mathcal V}_n$ in the range $[-1,1]$, inspired in part by the fact that the QCD axion potential has order 1 coefficients in this basis (see equation \eqref{eq:QCDcoeff}).

Figures \ref{fig:globalTuning1} and \ref{fig:globalTuning2} provide illustrative examples of some important qualitative features of the distribution of oscillon lifetimes. First, we observe islands of longevity, seen in figure \ref{fig:globalTuning1} as localized regions of exceptionally long lifetimes. In figure \ref{fig:globalTuning2}, this feature is manifested as plateaus in the distribution of lifetimes. We observe that each successive degree of freedom introduces a new longer-lived island of longevity which we observe to be exponentially more long-lived than the last. The probability of landing on one of these islands decreases with a scaling expected to be exponential in the number of degrees of freedom.

With these observations in mind, we introduce a notion of global tuning based on the cumulative probability of finding an oscillon at least as long-lived. Therefore, smaller values of this probability mean more extreme outliers, and thus higher degrees of global tuning. For example, according to our PQB simulations, summarized by figure \ref{fig:globalTuning2}, an oscillon of lifetime $\log_{10} mT = 3$ is tuned to one part in 2 (or 50\%). Oscillons of lifetime $\log_{10} mT = 7$ are tuned to one part in $8$ (or 10\%), and oscillons of lifetime $\log_{10} mT = 12$ are tuned to one part in 400 (or 0.2\%). Finally, the longest lived potential we observe in our search lives roughly $\log_{10} mT = 14$, and its tuning is roughly one part in 3000, or 0.03\%, although longer-lived oscillons may still reside further up the distributional tail.

\subsection{\label{sec:localTuning}Local tuning}

As opposed to global tuning, which deals with the statistics over large volumes of parameter space, local tuning attempts to quantify the sensitivity of an oscillon's lifetime to variations in its potential coefficients. If we understand the lifetime $mT$ as a function of the potential coefficients $\vec{\mathcal V}\equiv\{{\mathcal V}_1,\dots\}$, we can naturally introduce a local approximation of $mT$ as a function of its gradient and curvature, writing $\vec{\mathcal V} = \vec{\mathcal V}_0 + \delta \vec{\mathcal V}$, and $mT(\vec{\mathcal V}_0) = mT_0$,
\begin{equation}
\begin{aligned}
    G_i&=\f{\partial \log_{10}mT}{\partial {\mathcal V}_i} \,,\hspace{0.5cm}
    \mathbf{K}_{ij}=\f{\partial ^2\log_{10}mT}{\partial {\mathcal V}_i \partial {\mathcal V}_j}\,,\\
    \log_{10}mT &= \log_{10} mT_0 + \vec G\cdot \delta\vec{\mathcal V} + \f12 \delta\vec{\mathcal V}\cdot \mathbf{K}\cdot\delta\vec{\mathcal V}\,.
\end{aligned}
\end{equation}
In terms of this local approximation, we may quantify the sensitivity of the lifetime to local variations in $\vec{\mathcal V}$, as the minimum relative displacement of the potential coefficients $||\delta\vec{\mathcal V}||/||\vec{\mathcal V}_0||$ necessary to change the lifetime by an order of magnitude $\log_{10} mT_0\pm 1$. In other words, our local tuning metric is the solution to the following constrained optimization problem,
\begin{equation}
\begin{aligned}
    \te{minimize }&||\delta\vec{\mathcal V}||/||\vec{\mathcal V}_0||\,,\\
    \te{subject to }&\abs{\vec G\cdot \delta\vec{\mathcal V} + \f12 \delta\vec{\mathcal V}\cdot \mathbf{K}\cdot\delta\vec{\mathcal V}}>1\,.
\end{aligned}
\end{equation}
We denote the minimal value $\nu \equiv ||\delta\vec{\mathcal V}||/||\vec{\mathcal V}_0||$.

Consider the potential $\vec{\mathcal V} = (1,1/2,-1)$, for which an oscillon lives approximately $\log_{10}mT = 14$. Using the above measure of tuning, and a grid based approximation to the gradient and Hessian, we calculate $\nu \approx 0.03$. In other words, a 3\% variation in the potential parameters corresponds to an order of magnitude change in the lifetime of the oscillon. This is substantially \emph{less} tuned than one would expect from our global metric, in which this potential is $0.03\%$ tuned. This is a reflection of the structure of the lifetime landscape, which contains islands of stability seen in figures \ref{fig:landscape1}, \ref{fig:globalTuning1} and \ref{fig:globalTuning2}.

\section{\label{sec:illustrativeExamples}Illustrative examples}

Here we apply our framework to a series of potentials that have been studied extensively in previous literature, with the aim to reproduce and expand upon known results. The main goal is to show how our methods can accommodate a wide variety of potentials: both with or without parity, and with or without periodicity. We compare the results of our PQB framework to explicit numerical simulation. When simulating the equations of motion explicitly (as in appendix \ref{sec:simulation}), the wall-clock time is at least proportional to the oscillon lifetime, which becomes computationally prohibitive for lifetimes beyond $10^{10}/m$. Our framework can bypass this scaling since time has been explicitly integrated out, allowing us to predict the existence of very long-lived oscillons, well in excess of the lifetimes we can simulate explicitly.

The results of this section are presented in the form of ``power vs frequency'' curves, which represent the instantaneous flux radiated by the oscillon at a particular fundamental frequency $\omega$. The oscillon fundamental frequency $\omega$ monotonically increases with time, and can therefore be thought of as a time coordinate. For a detailed review of how to interpret these plots and their features, see figure \ref{fig:genericPower}. 
\subsection{\label{sec:monodromy}Axion monodromy}
\begin{figure}
    \centering
    \includegraphics[width = \columnwidth]{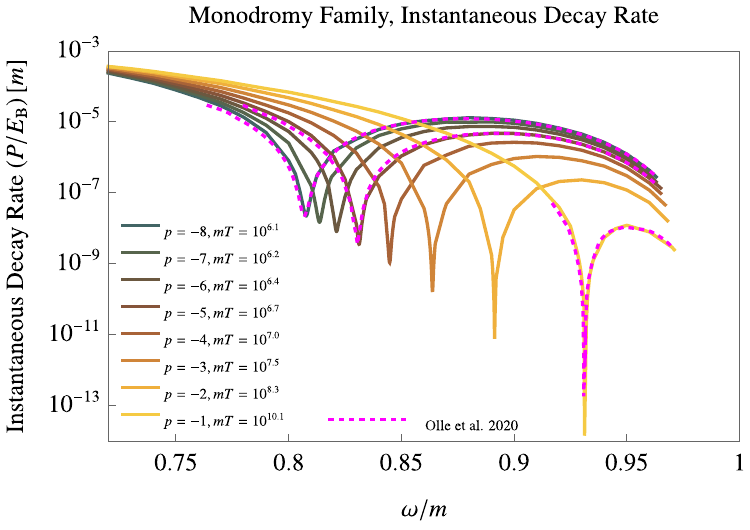}
    \caption{The instantaneous decay rate $P/E_\te{B}$ of the oscillons in the monodromy potentials \eqref{eq:monodromyPotential} for $p = -1,\dots,-8$, calculated in the PQB formalism, versus the results of Olle et al. \cite{olle2020recipes}. Here, the power $P$ and binding energy $E_\te{B}$ are computed as in section \ref{sec:physicalQuasibreatherMinTech}. As $p$ scans from $-8$ to $-1$, the third harmonic dip migrates to larger frequencies where the fifth harmonic is further suppressed by geometric decoupling, leading to increased lifetime.
    To obtain the PQB results, we start with a two-non-perturbative harmonic approximation and used three non-perturbative harmonics to obtain better accuracy near the dips. At frequencies below the dip frequency, we see a small shift in the PQB formalism vs \cite{olle2020recipes}, which may arise from the need to use more non-perturbative harmonics at lower frequencies or because the Fourier series representation of \eqref{eq:monodromyPotential} converges slowly.}
    \label{fig:monodromy}
\end{figure}
Monodromy refers to the non-trivial winding of an axionic degree of freedom, which effectively endows the axion with an aperiodic potential at low energies \cite{silverstein2008monodromy,dubovsky2012axion,arvanitaki2010string}. In general, the resulting monodromy potentials share the property that they asymptote to a power law at large field values. 
A common family of potentials which interpolate between the asymptotic power-law and the small-field mass $m$ is
\begin{align}
\label{eq:monodromyPotential}
    V(\phi) = \f{m^2 f^2}{p}\p{\ps{\p{\f{\phi}{f}}^2 + 1}^{p/2} - 1}\,,
\end{align}
where $p$ scans the asymptotic power-law. The potential \eqref{eq:monodromyPotential}
has been widely studied, and has been shown to support very long-lived oscillons, in excess of $10^{10}$ cycles \cite{amin2012oscillons,salmi2012radiation,zhang2020classical,olle2020recipes}. 

In figure \ref{fig:monodromy}, we summarize our study of the oscillon life-cycles as we scan $p$ from $-8$ to $-1$. In general, we find good agreement with the results of \cite{olle2020recipes}: the power versus frequency curves and lifetimes broadly match the predictions of \cite{olle2020recipes} in the cases we have mutually studied $p = -8,-5,-4,-1$ although there are minor differences.

As $p$ increases from $-8$ to $-1$, the lifetime of the corresponding oscillon increases dramatically, from $10^6$ to $10^{10}$ cycles. This is due to the simultaneous action of the two longevity mechanisms identified in section \ref{sec:lifecycle}. Specifically, the third harmonic experiences totally destructive interference at an exceptional frequency that is larger with increasing $p$. Therefore, as $p$ grows, the third harmonic dip moves deeper into the frequencies where geometric decoupling dominates, which further suppresses the fifth harmonic.

A natural conjecture is that the oscillons of \eqref{eq:monodromyPotential} are unusually long-lived because of the asymptotic power law in the potential. However, our results in figure \ref{fig:monodromy} indicate that longevity is dominated by large frequencies, where field amplitudes are too small for the asymptotic behavior to take over. In particular, for $p = -1$, the field amplitude at the origin is roughly $\phi(0)\approx 1.5 f$, far too small to be sensitive to the flatness of the potential at large $\phi/f$. Therefore, we conjecture that it is \emph{not} the asymptotic form, but the details of the connection between $\f12m^2\phi^2$ and $\phi^p$ that determine the oscillon lifetime.

In the oscillon literature, many examples of extremely long-lived oscillons are obtained with monodromy potentials. Thus, a natural question is whether all monodromy potentials share a common feature leading to longevity, or whether simple examples such as \eqref{eq:monodromyPotential} happen to live in a tuned island of longevity. In the language we introduced in section \ref{sec:globalTuning}, in order to quantify the link between monodromy and longevity we would need to know the probability distribution from which monodromy potentials are chosen. In the absence of this non-trivial construction, we are left with a case-by-case analysis of particular potentials, which, in this probabilistic view, may suffer from sampling bias.

\subsection{$\phi^4$ theory}
\begin{figure}
    \centering
    \includegraphics[width = \columnwidth]{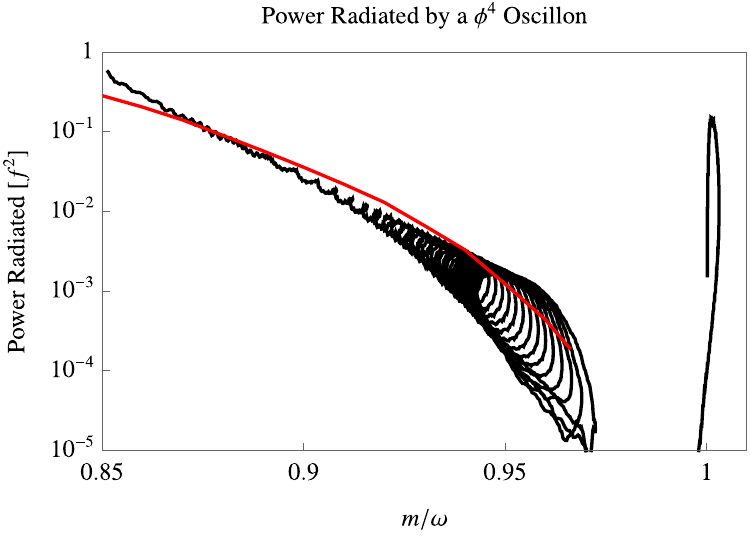}
    \caption{Comparison of the explicitly simulated $\phi^4$ oscillon (black) with the physical quasibreather trajectory (red) truncated to the leading three harmonics $C_0,S_1,C_2$ (in the notation of appendix \ref{sec:ThePhysicalQuasibreatherFormalism}), all treated non-perturbatively. The oscillating behavior is a symptom of linear instability, although crucially, it does not destroy the oscillon, since the size of the oscillations is controlled by nonlinearity. For technical details of the explicit simulation, see appendix \ref{sec:simulation}.}
    \label{fig:phi4Power}
\end{figure}
$\phi^4$ theory is the quintessential example of spontaneously broken parity symmetry. It is well known to host moderately long-lived oscillons, which have been studied in previous work \cite{honda2002fine,fodor2006oscillons,andersen2012four,salmi2012radiation,gleiser2019resonant}. Shifting to the broken vacuum and fixing the mass of $\phi = \theta/f$ to be $m$, we arrive at the following parity-violating potential
\begin{align}
    V(\theta) = m^2 f^2\p{\f12\theta^2 - \f{1}{2}\theta^3 + \f{1}{8}\theta^4}\,.
\end{align}
In order to properly compute the physical quasibreather in this potential, the first three harmonics $C_0,S_1,C_2$ must be treated non-perturbatively. As is evident from the numerical simulation (figure \ref{fig:phi4Power}), the $\phi^4$ physical quasibreathers are linearly unstable over the entire range of $\omega$ for which the oscillon is long lived. The instability that occurs at linear order in the PQB background is, however, stabilized by self-interaction, leading to quasiperiodic oscillations. These nonlinear oscillations are visible as dense curly-Q's in the Power vs Frequency plot (figure \ref{fig:phi4Power}).

Our explicit numerical simulation yields an approximate lifetime of $6000/m$, which is close to the PQB prediction of $5900/m$. This confirms the earlier results in \cite{fodor2006oscillons}.

\subsection{The QCD axion}
\begin{figure}
    \centering
    \includegraphics[width = \columnwidth]{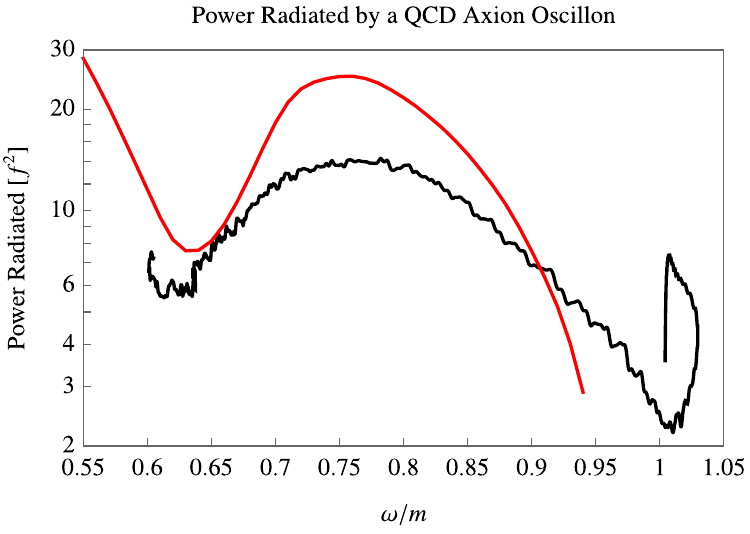}
    \caption{Comparison of the explicitly simulated QCD axion oscillon (black) to the PQB formalism (red) with three non-perturbative harmonics. The radiated power is so large that the orthogonal deformation is non-perturbative, leading to disagreement within a factor of a few, although the shape of the physical quasibreather curve still captures the qualitative features of the simulated oscillon. Namely, it shows that there is a dip around $\omega = 0.6 m$ where the fifth harmonic is dominant. This region, in which the third harmonic is confined and non-perturbatively large, constitutes most of the oscillon's lifetime.}
    \label{fig:QCD}
\end{figure}

The QCD axion is the best studied example of a scalar field described by a periodic potential, and could allow for oscillons with cosmological observables. At low temperatures, the QCD axion potential is dominated by strong dynamics, giving rise to the potential \cite{di2016qcd,visinelli2018dilute},
\begin{align}
\label{eq:QCDPotential}
    V(\phi) = -m_\pi^2 f_\pi^2\sqrt{1 - \f{4 m_u m_d}{(m_u + m_d)^2}\sin^2\p{\f{\phi}{2 f}}}\,.
\end{align}
A simple Taylor expansion about $\theta = 0$ reveals that \eqref{eq:QCDPotential} has a smaller quartic term than the simple cosine potential, which leads us to expect the oscillons in \eqref{eq:QCDPotential} to be shorter-lived than the sine-Gordon oscillon. Indeed, this expectation is confirmed by the physical quasibreather framework and explicit numerical simulation (see figure \ref{fig:QCD} and appendix \ref{sec:simulation}).

In order to compute the lifetime in the physical quasibreather formalism, we calculate the potential's Fourier coefficients
\begin{align}
\label{eq:QCDcoeff}
    \vec{\mathcal V} = \{1.427, -0.648, 0.336,\dots\}
\end{align}
Our formalism can accommodate many non-zero Fourier coefficients, although only the first three (and a fourth to normalize the mass) are necessary to converge within 1\% of the true potential; adding more terms has a negligible impact on the results of explicit simulation and on the result of our PQB formalism.

The result of this analysis is that the QCD axion oscillon is relatively short-lived compared to the average oscillon, living about $400/m$. However, it is interesting to observe that this oscillon spends most of its life with a confined third harmonic, undergoing very large central-amplitude oscillations of order $15 f$. Although the short lifespan of the QCD axion oscillon means that it will only leave its cosmological imprint shortly after formation, the large amplitude and violent deaths of these oscillons may have observational implications.

\section{Conclusion}
Real scalar fields play a central role in many theories of early universe cosmology and dark matter. Many of these theories predict attractive self-interactions that allow the scalars to form quasistable oscillons. Understanding oscillon lifetime is necessary for determining whether oscillons only play a role in early universe cosmology, or whether they may also survive until the present day and lead to dramatic astrophysical signatures.

In this work, we have expanded the quasibreather approximation into a formalism for computing the properties of oscillons that naturally incorporates realistic boundary conditions. We defined the physical quasibreather by finding initial conditions of the nonlinear wave equation that simultaneously obey radiative boundary conditions \emph{and} specify a quasibreather solution. As the closest quasibreather to a physical oscillon, the PQB provides a raw approximation for the oscillon profile (see e.g. figure \ref{fig:DPQBAnatomy}) which is increasingly accurate in the limit of long lifetimes. Furthermore, the PQB represents the solutions to which the oscillon is instantaneously and locally attracted to during its evolution. When understood as a stable perturbation to a PQB, the oscillon borrows its properties from its PQB partner, including its radial profile and radiation rate \footnote{Although, see appendix \ref{sec:ThePhysicalQuasibreatherFormalism} for a full explanation of the sense in which the oscillon is a stable perturbation of a PQB}. When the PQB becomes linearly unstable, nonlinear quasiperiodic oscillations emerge, whose size is controlled by higher-order terms, as depicted in figures \ref{fig:phi4Power}, \ref{fig:SGLinStab}, and \ref{fig:SGEvo}. In other words, linear instability often does not result in the death of the oscillon. Further, we have demonstrated that the PQB and the minimum radiation quasibreather differ at 6th order in the background, explaining the success of the `minimum radiation quasibreather ansatz.'

Since the PQB offers an accurate description of the oscillon structure, we have used it to understand the oscillon's form-factor and the resulting mechanisms which control longevity. Specifically, as the oscillon radiates its energy away, its central amplitude decreases, causing self-interactions to become weaker; as a result, the oscillon becomes much larger than the radiation wavelengths, suppressing the radiated power (see figure \ref{fig:genericPower}). At these high frequencies $(\omega\to m)$, the large oscillon core naturally leads to the rapid self-interference of radiation. When the self-interference is destructive, the total radiation is suppressed by another power of the form-factor. While both these effects, geometric decoupling and destructive interference, are generic features of oscillon evolution, the longest-lived oscillons are a consequence of these two effects occurring simultaneously at large oscillon frequencies close to $m$ (as in figure \ref{fig:monodromy}). Finally, we have understood the physics of oscillon death as a further consequence of weak self-interaction: past a certain critical frequency the energy of the PQB is forced to increase, and the oscillon cannot sustain its proximity to its PQB partner. Using our understanding of the mechanisms responsible for oscillon longevity and death, we have constructed the family of \emph{frustrated quadratic} potentials which support extremely long-lived oscillons, living more than $10^{18}$ cycles (see section \ref{sec:prescription}).

There are several computational advantages provided by our methodology. First, the oscillon evolution is computed in a time-independent way, separating the physical lifetime of the oscillon from the computational wall-clock time it takes to evolve numerically. Second, our formalism naturally incorporates non-perturbative harmonics, and potentials without even parity. Third, all perturbative harmonics may be efficiently computed by taking advantage of sparse linear algebra. Fourth, we have introduced the Fourier basis as the natural basis for expanding potentials when studying oscillons. In this basis, the form of the mode equations is especially simple, allowing us to write down analytical expansions for the mode potentials that converge everywhere. Fifth, our formalism provides a natural language for studying the stability of oscillons. Finally, the speed of our numerical techniques has enabled us to study extremely long-lived oscillons, and has yielded the first prediction of cosmologically long lived periodic potentials (see figure \ref{fig:frustrationLifetime}).

Using our efficient numerical techniques, we scan over degrees of freedom in axionic potentials (see figures \ref{fig:landscape1} and \ref{fig:globalTuning1}), allowing us to probe the genericity of long-lived oscillons (see figure \ref{fig:globalTuning2}). Important outcomes of this parameter scan include the identification of features in the lifetime landscape with the mechanisms of longevity (see section \ref{sec:lifecycle}), and the realization that extremal lifetimes may scale at least exponentially in the number of potential degrees of freedom. At the same time, long lifetimes are not particularly fine-tuned, since as few as three degrees of freedom are enough to generate oscillons that survive until last-scattering ($4\times 10^{6}$ cycles) with only $15\%$ global tuning (as defined in section \ref{sec:globalTuning}), and oscillons that live until today ($10^{11}$ cycles) with 0.5\% global tuning.

\begin{acknowledgments}
We thank Savas Dimopoulos, Asimina Arvanitaki, Sebastian Baum, Robert Lasenby, Horng-Sheng Chia, Anirudh Prabhu, and Dmitriy Zhigunov for stimulating conversations and helpful comments on the manuscript. D.C. is grateful for the support of the National Science Foundation under grant number PHY-1720397, and for the support of the SITP. T.G.T. is grateful for the support of the Department of Energy under grant number DE-SC0020266.
Some of the computing for this project was performed on the Sherlock and Farmshare clusters. We would like to thank Stanford University and the Stanford Research Computing Center for providing computational resources and support that contributed to these research results.
\end{acknowledgments}

\appendix

\section{\label{sec:ThePhysicalQuasibreatherFormalism}The physical quasibreather formalism}

In this appendix, we outline the precise definition of the physical quasibreather (PQB) and its orthogonal deformation (OD) introduced in section \ref{sec:physicalQuasibreatherMinTech}. We will see that the orthogonally deformed PQB is an instantaneous solution of the equations of motion that satisfies outgoing boundary conditions. We then find oscillonic solutions of the equations of motion that are perturbations of the orthogonally deformed PQB. By studying the evolution and stability of these perturbations, we arrive at a sense in which the orthogonally deformed PQB can be an attractor, which we apply to study oscillon stability in appendix \ref{sec:FloquetAnalysis}.

\subsection{Quasibreathers}

Physical potentials may be interpreted in terms of $n$-particle interactions, and therefore possess Taylor expansions around their vacua. Consequently, a periodic field configuration with fundamental frequency $\omega$ will only couple to modes oscillating with integer multiples of this fundamental frequency. In other words, physical non-linear wave equations possess periodic orbits, which may be interpreted as a Fourier series in time. This is in contrast to unphysical potentials that may not be interpreted in terms of integer-number particle interactions, which can at best possess quasiperiodic orbits.

For the remainder of the appendix, we move into dimensionless units with $m = f = 1$. The non-linear wave equation for the field $\theta$ in a potential $V$ is then
\begin{align}
\label{eq:EOMgeneral}
    0&=\ddot\theta - \nabla^2\theta + V'(\theta)\,.
\end{align}
As we have argued above, $V$ must possess a Taylor series, and therefore $\theta$ may be expanded as a series of integer harmonics
\begin{align}
\label{eq:periodic}
    \theta = \sum_{n\in\N_0}S_n( r,\omega)\sin (n\omega t + \delta_n)\,,
\end{align}
where $\delta_n$ is a phase, and we have taken spherical symmetry for simplicity. Without loss of generality, we may take $\delta_1 = 0$. We say that a solution of the form \eqref{eq:periodic} is generated by the frequency $\omega$ if $S_1$ is non-zero, and the only non-zero higher harmonics $S_n$ are those that couple to $S_1$, consistent with closure of the equations of motion. We then define the quasibreather as the solution generated by $\omega$.

Using this definition, we may compute the generic form of the quasibreather. Consider the generic potential
\begin{align}
    V(\theta) = \f12\theta^2 + \f13\lambda_3\theta^3 + \f14\lambda_4\theta^4 +\dots\,.
\end{align}
From the symmetries of sine and cosine, we observe that
\begin{align}
\notag
    (\sin\omega t)^n &=\left\{\begin{array}{cc}\sum_k a_k\sin n_k\omega t,\,n_k\in\N_\te{odd},&n\te{ is odd,}\\\sum_k b_k\cos m_k\omega t,\,m_k\in\N_\te{even},&n\te{ is even,}\end{array}\right. \\\notag
    (\cos\omega t)^n &= \left\{\begin{array}{cc}\sum_k c_k\cos n_k\omega t,\,n_k\in\N_\te{odd},&n\te{ is odd,}\\\sum_k d_k\cos m_k\omega t,\,m_k\in\N_\te{even},&n\te{ is even.}\end{array}\right.
\end{align}
The case of parity $V(\theta)= V(-\theta)$ offers a pleasant simplification, decopling the even harmonics and the odd harmonics from one another. Thus, potentials with parity have quasibreathers of the form
\begin{align}
\label{eq:QBparity}
    \theta_\te{QB} = \sum_{n\in\N_\te{odd}}S_n( r,\omega)\sin n\omega t\,,
\end{align}
and a periodic solution of the form $\theta = \sum_{n\in\N_\te{even}}S_n( r,\omega)\sin n\omega t\,,$
although it is not a quasibreather because it is not generated by $\omega$.
Quasibreathers in potentials without parity possess expansions
\begin{align}
\label{eq:QB}
    \theta_\te{QB} = \sum_{n\in\N_\te{odd}}S_n( r,\omega)\sin n\omega t + \sum_{n\in\N_\te{even}}C_n( r,\omega)\cos n\omega t\,,
\end{align}
where $\N_\te{even}$ contains 0. Thus, we have identified the form of the quasibreathers of the non-linear wave equation when $V$ represents a physical interaction.

Inserting the form \eqref{eq:QB} into \eqref{eq:EOMgeneral}, one arrives at the set of mode equations
\begin{align}
\label{eq:ModeQB}
\notag
    0&=-(n\omega)^2C_n-C_n'' - \f{d-1}{r}C_n' + V_n'(C,S)\,,n\in \N_\te{even}\,,\\
    0&=-(n\omega)^2S_n -S_n'' - \f{d-1}{r}S_n' + V_n'(C,S)\,,n\in \N_\te{odd}\,,
\end{align}
where $d$ is number of spatial dimensions, and $V_n$ is defined through the equation
\begin{align}
    V'(\theta_\te{QB}) &= \sum_{n\in\N_\te{odd}} V_n(C,S)\sin n\omega t + \sum_{n\in\N_\te{even}} V_n(C,S)\cos n\omega t\,.
\end{align}
Equation \eqref{eq:ModeQB} is a system of second order ordinary differential equations, and therefore each degree of freedom $S_n,C_n$ must be constrained by two boundary conditions.

In order to discuss boundary conditions, we define the number $n_0$ as the least integer such that $n_0\omega > 1$, so that bound harmonics have $n< n_0$ and radiative harmonics have $n\geq n_0$. Regularity at the origin places a non-trivial constraint on all harmonics, that all $S_n$ and $C_n$ must have zero first derivative at the origin. However, regularity at spatial infinity is only a constraint on the bound modes, $n < n_0$; all radiative harmonics decay geometrically as they propagate to spatial infinity. Thus, for a quasibreather, the radiative harmonics are only constrained by regularity at the origin, and the space of possible quasibreathers has dimension equal to the number of radiative modes. In other words, one has the freedom to pick the amplitude of the radiative modes at the origin, and the result will still be a quasibreather. The authors of \cite{fodor2019review} have alleviated this ambiguity by picking a specific quasibreather out of this manifold: the \emph{minimum radiation quasibreather}, whose radiative tails are the smallest. Instead, we pick the physical quasibreather (PQB), defined below, which is perturbatively close to a radiating solution.

\subsection{The deformed mode equations}
A localized field configuration with a finite lifetime necessarily radiates its energy to spatial infinity, and therefore satisfies radiative boundary conditions at spatial infinity. 
In this section, we introduce the concept of the \emph{physical quasibreather} (PQB), which is, in a precise sense, the quasibreather closest to a physical configuration satisfying radiative boundary conditions.

First, we define the \emph{orthogonal deformation} (OD) of the quasibreather \eqref{eq:QB}, which consists of adding $90^\circ$ out-of-phase components $s_n$ and $c_n$ to the radiative harmonics in order to satisfy radiative boundary conditions
\begin{equation}
\begin{aligned}
\label{eq:QBOrthogonal}
    \theta &=\sum_{n\in\N_\te{odd}}S_n(t, r,\omega)\sin n\omega t + \sum_{n\in\N_\te{even}}C_n(t, r,\omega)\cos n\omega t\\
    &\hspace{0.35cm}\sum_{n\in\N_\te{even}^{\geq n_0}}s_n(t, r,\omega)\sin n\omega t + \sum_{n\in\N_\te{odd}^{\geq n_0}}c_n(t, r,\omega)\cos n\omega t\,.
\end{aligned}
\end{equation}
Notice that here we've introduced a time dependence to the modes, which accounts for the fact that a radiating solution cannot have a stationary profile. This formulation will be useful for studying initial conditions of interest, namely, those which specify a quasibreather and orthogonal deformation that together satisfy outgoing boundary conditions. 
Although \eqref{eq:QBOrthogonal} is a vast overparametrization of a single field, we recognize that a solution of the \emph{deformed mode equations} \eqref{eq:ModeEQ}
\begin{align}
    \label{eq:ModeEQ}
    \notag
    0&=\ddot S_n - 2 n \omega \dot c_n - (n\omega)^2S_n - S_n'' - \f{d-1}{r}S_n' + V_{S_n}'\,,\\\notag
    0&=\ddot c_n + 2 n \omega \dot S_n - (n\omega)^2c_n - c_n'' - \f{d-1}{r}c_n' + V_{c_n}'\,,\\\notag
    0&=\ddot C_n + 2 n \omega \dot s_n - (n\omega)^2C_n - C_n'' - \f{d-1}{r}C_n' + V_{C_n}'\,,\\
    0&=\ddot s_n - 2 n \omega \dot C_n - (n\omega)^2s_n - s_n'' - \f{d-1}{r}s_n' + V_{s_n}'\,,
\end{align}
is also a solution of the full equation of motion \eqref{eq:EOMgeneral}. Equation \eqref{eq:ModeEQ} is obtained from the equations of motion \eqref{eq:EOMgeneral} by substituting \eqref{eq:QBOrthogonal} and collecting the terms proportional to $\sin n\omega t$ or $\cos n\omega t$ for a given $n$, setting them independently to zero. Intuitively, when the time dependence of the harmonic functions $S,C,s,c$ is slow, they have the usual interpretation as the profiles of quasistationary modes.  Here the functions $V_X' = V_X'(S_n,C_n,s_n,c_n)$, are the \emph{mode potentials}, in which we have suppressed functional dependence for brevity. The mode potentials are defined by the equation
\begin{equation}
\begin{aligned}
    V'(\theta) &= \sum_{n\in\N_\te{odd}}V_{S_n}'\sin n\omega t + \sum_{n\in\N_\te{even}}V_{C_n}'\cos n\omega t\\
    &\hspace{0.35cm}\sum_{n\in\N_\te{even}}V_{s_n}'\sin n\omega t + \sum_{n\in\N_\te{odd}}V_{c_n}'\cos n\omega t\,,
\end{aligned}
\end{equation}
where $\theta$ is written in the form of equation \eqref{eq:QBOrthogonal}, and $V_{S_n}',V_{C_n}',V_{s_n}',V_{c_n}'$ are pure functions of $S_n,C_n,s_n,c_n$.

During the evolution of equation \eqref{eq:ModeEQ}, $\omega$ is treated as a constant. This is not in contradiction to the usual understanding that the fundamental frequency of the oscillon increases with time. For the purpose of the mode equations \eqref{eq:ModeEQ}, $\omega$ is understood as a choice of a fixed parameter, independent of the time variation of the modes $S,C,s,c$ themselves. For certain initial conditions, and for certain choices of $\omega$, there will be periods of time over which $S,C,s,c$ vary slowly, and it is during these periods that $\omega$ may be interpreted as the instantaneous frequency of the oscillon.

In other words, there is no a priori reason to choose a particular $\omega$ for a particular field configuration, and one may only think of $\omega$ as an instantaneous frequency in the context of certain initial conditions. Thus, the following paragraphs are dedicated to specifying initial conditions which allow $\omega$ to be interpreted as the instantaneous frequency of an oscillon, where the oscillon is perturbatively close to a quasibreather. The smaller the orthogonal deformation, the better this interpretation is, and the longer it holds. In this sense, $\omega$ may be conceptualized as an adiabatic parameter, although one should not confuse it with an externally controlled parameter --- in our framework, it is a constant that parametrizes the decomposition  \eqref{eq:QBOrthogonal} of solutions to \eqref{eq:EOMgeneral}.

We now specify the following consistent set of initial and boundary conditions, in which we treat $s_n$ and $c_n$ as linear perturbations. Here, we take $V_X' = V_X'(S_n(0,r),C_n(0,r),0,0)$, and we define $\delta V_X' \equiv \sum_{n\geq n_0} s_n(0,r)\partial_{s_n} V_X'+ c_n(0,r)\partial_{c_n} V_X'$ (note the absence of a constant term in $\delta V_X'$ is a consequence of (a), below). A complete and consistent set of initial and boundary conditions associated with \eqref{eq:ModeEQ}, that exactly specify a quasibreather and orthogonal deformation at $t = 0$ is
\begin{align*}
\te{(a)}&\textbf{ Initial Quasibreather:}\\
    0&=-(n\omega)^2S_n(0,r) - S_n''(0,r) - \f{d-1}{r}S_n'(0,r) + V_{S_n}'\,,\\
    0&=-(n\omega)^2C_n(0,r) - C_n''(0,r) - \f{d-1}{r}C_n'(0,r) + V_{C_n}'\,,\\
\te{(b)}&\textbf{ Initial Deformation:}\\
    0&=- (n\omega)^2c_n(0,r) - c_n''(0,r) - \f{d-1}{r}c_n'(0,r) + \delta V_{c_n}'\,,\\
    0&=- (n\omega)^2s_n(0,r) - s_n''(0,r) - \f{d-1}{r}s_n'(0,r) + \delta V_{s_n}'\,,\\
\te{(c)}&\textbf{ Maximally stationary:}\\
    0&=\dot S_{n\geq n_0}(0,r)=+2n\omega \dot S_{n < n_0}(0,r) +  \delta V_{c_n}'\,,\\
    0& = \dot C_{n\geq n_0}(0,r)=-2n\omega \dot C_{n < n_0}(0,r) +  \delta V_{s_n}'\,,\\
    0&=\dot s_{n}(0,r) = \dot c_{n}(0,r)\,,\\
\te{(d)}&\textbf{ Regularity:}\\
0&=S_n'(t,0)=C_n'(t,0)=s_n'(t,0)=c_n'(t,0)\,,\\
0&=S_{n<n_0}'(t,\infty)=C_{n<n_0}'(t,\infty)\,,\\
\te{(e)}&\textbf{ Radiative:}\\
0&=\lim_{r\to\infty} r^{\f{1-d}{2}}(r^{\f{d-1}{2}}c_n(t,r))' + \sqrt{(n\omega)^2 - 1}S_n(t,r)\,,\\
0&=\lim_{r\to\infty} r^{\f{1-d}{2}}(r^{\f{d-1}{2}}S_n(t,r))'- \sqrt{(n\omega)^2 - 1}c_n(t,r)\,,\\
0&=\lim_{r\to\infty} r^{\f{1-d}{2}}(r^{\f{d-1}{2}}C_n(t,r))' + \sqrt{(n\omega)^2 - 1}s_n(t,r)\,,\\
0&=\lim_{r\to\infty} r^{\f{1-d}{2}}(r^{\f{d-1}{2}}s_n(t,r))'- \sqrt{(n\omega)^2 - 1}C_n(t,r)\,.
\end{align*}
Our initial condition (a) selects $S_n$ and $C_n$ which specify a quasibreather. This quasibreather is one which may be orthogonally deformed to satisfy radiative boundary conditions, and it is this quasibreather which we call the PQB.

Because we have broken the time translation symmetry of the quasibreather by satisfying radiative boundary conditions, the modes $S,C,s,c$ are endowed with an irreducible time dependence. The maximally stationary condition (c) shows that this time dependence is proportional to the pointwise small deformations $s$ and $c$. Since $s$ and $c$ obey a homogeneous system of equations (b), their amplitude everywhere must uniformly go to zero as their amplitude at $r = \infty$ goes to zero. From (e), we see that $c_n\propto S_n$ and $s_n\propto C_n$ at spatial infinity. Thus we conclude that if $S_n$ and $C_n$ possess small radiative tails, then $s_n$ and $c_n$ become pointwise small everywhere, and the time variation of the modes uniformly approaches zero. This is the limit in which the oscillon is long-lived, and approaches the quasibreather; in this same limit, the interval over which this approximation is valid, during which $\omega$ may be thought of as an instantaneous frequency, becomes longer.

\subsection{The asymptotic attractor}

\begin{figure}
    \centering
    \includegraphics[width = \columnwidth]{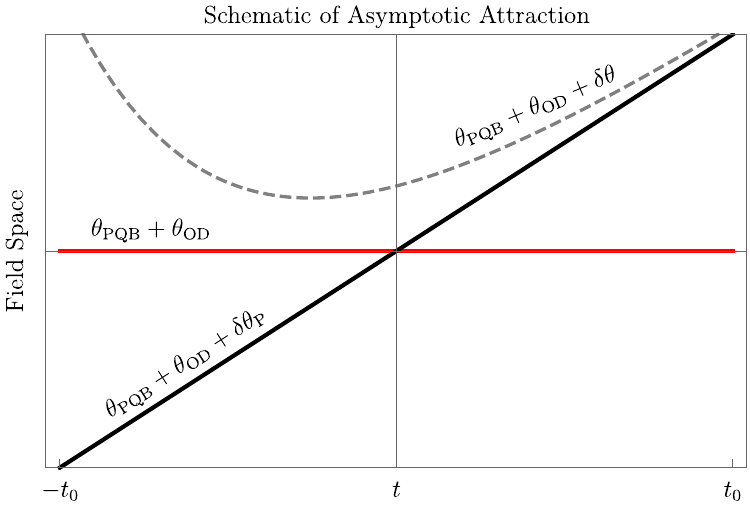}
    \caption{The asymptotic attractor (red) is approached as the inhomogeneous solution goes to zero. The homogeneous terms, representing the initial conditions at $t = -t_0$ cannot converge exactly to zero by the time the inhomogeneous solution passes through zero, and therefore the perturbation never exactly reaches the asymptotic attractor.}
    \label{fig:AA}
\end{figure}

\begin{figure}
    \centering
    \includegraphics[width = \columnwidth]{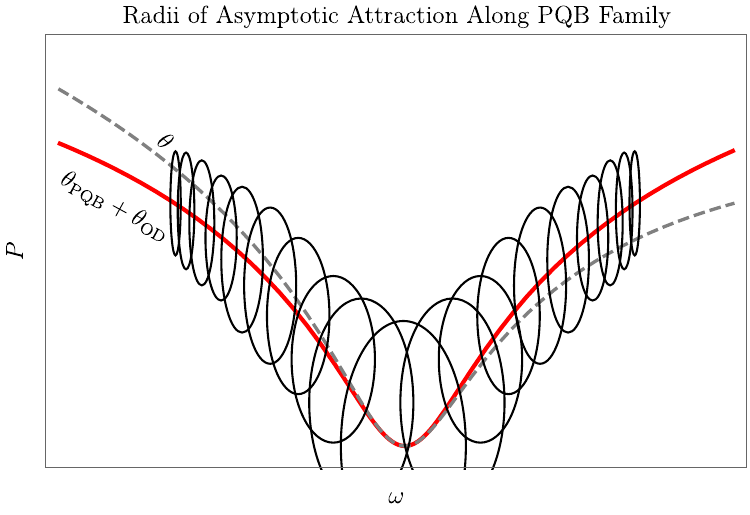}
    \caption{Here we have a schematic Power Radiated (as a proxy for field-space) vs Oscillon Frequency plot for the family of deformed PQB (red) and an oscillon trajectory (dashed grey). Each ellipse centered on a deformed PQB represents the domain of frequencies and field values over which that specific quasibreather is an asymptotic attractor. As the oscillon trajectory enters an attractive region, it moves closer to the attractive deformed PQB. Consequently, it is also drawn into the attractive vicinity of the neighboring PQBs. Therefore, the oscillon is forced to approach the red trajectory as the radii of attraction get larger and larger towards the bottom of the dip. After traversing the dip, the deformed PQB radii of attraction begin to shrink, and the oscillon trajectory begins to diverge from the deformed PQB trajectory. In this latter half of the evolution, we see how the deformed PQB trajectory does not act as a standard attractor, but can still be described as an asymptotic attractor. To see this, notice how the oscillon instantaneously moves closer to the quasibreather when entering each new attractive bubble.}
    \label{fig:AE}
\end{figure}

The initial conditions (a-e) specify a solution to the nonlinear wave equation that, at $t = 0$ is exactly an orthogonally deformed physical quasibreather (to linear order in the deformation). More practically, we want to understand the evolution of the oscillon in the neighborhood of this deformed physical quasibreather, before and after this particular point. To this end, we introduce the perturbation $\delta\theta(t,r)$, which simultaneously absorbs the time dependence of the modes in (a-e), and deviations from the orthogonally deformed physical quasibreather.
Specifically, a field configuration $\theta$ describing a physical oscillon can be expanded as
\begin{align}
\label{eq:decomposition}
    \theta = \theta_\te{PQB} + \theta_\te{OD} + \delta\theta\,,
\end{align}
with
\begin{align}
    \theta_\te{PQB}&=\sum_{n\in\N_\te{odd}}S_n(0, r,\omega)\sin n\omega t + \sum_{n\in\N_\te{even}}C_n(0, r,\omega)\cos n\omega t\,,\\
    \theta_\te{OD}&=\sum_{n\in\N_\te{even}^{\geq n_0}}s_n(0, r,\omega)\sin n\omega t + \sum_{n\in\N_\te{odd}^{\geq n_0}}c_n(0, r,\omega)\cos n\omega t\,.
\end{align}
Crucially, $\theta_\te{PQB}$ is exactly a quasibreather solution, and $\theta_\te{OD}$ is exactly periodic in time, as opposed to $\delta\theta$, which characterizes the secular evolution of the oscillon in the vicinity of the physical quasibreather at $\omega$.
Inserting \eqref{eq:decomposition} into \eqref{eq:EOMgeneral}, we arrive at the following equation for $\delta\theta$ at linear order,
\begin{align}
\label{eq:perturbationEQ}
\notag
0&=\delta\ddot\theta - \nabla^2\delta\theta+ V''(\theta_\te{PQB})\delta\theta\\& + \sum_{n < n_0} (\delta V_{c_n}'\cos n\omega t + \delta V_{s_n}'\sin n\omega t)\,.
\end{align}
This is a sourced equation, representing the fact that the physical quasibreather with orthogonal deformation does not conserve energy on its own. As a linear equation, $\delta\theta$ may be decomposed into a sum of homogeneous terms, which obey the homogeneous equation
\begin{align}
\label{eq:perturbationHomEQ}
0&=\delta\ddot\theta_H - \nabla^2\delta\theta_H+ V''(\theta_\te{PQB})\delta\theta_H\,,
\end{align}
and one particular solution $\delta\theta_P$, that obeys the sourced equation \eqref{eq:perturbationEQ}, which we take to be identically zero at $t = 0$. In the absence of homogeneous terms, it is this particular solution $\delta\theta_P$ which satisfies the initial conditions (a-e). Therefore, the homogeneous terms represent perturbations around those initial conditions. If the homogeneous solutions of \eqref{eq:perturbationEQ} are stable, then we say that $\theta_\te{PQB} + \theta_\te{OD}$ is an \emph{asymptotic attractor}. 

The usefulness of the construction $\delta\theta$ is that it contains all information about the linear stability of the oscillon \footnote{This is an oversimplification, since equation \eqref{eq:perturbationHomEQ} cannot be stable in the sense that it only has decaying modes. Stability will emerge out of nonlinear corrections, but for the sake of a simple discussion, we save this technicality for appendix \ref{sec:FloquetAnalysis}.}. Just like in a standard damped oscillator, linear stability represents an exponential approach to the inhomogeneous solution. In other words, it is enough to study the stability of the homogeneous equation \eqref{eq:perturbationHomEQ} with the tools of Floquet theory. The full picture of how the one-parameter family of deformed physical quasibreathers, parametrized by the frequency $\omega$, acts like an attractor may be understood in the following picture. Before $t = 0$, the particular solution $\delta\theta_P$ is approaching 0 (see Figure \ref{fig:AA}). If the homogeneous terms are stable, then the field $\theta$ is approaching the deformed physical quasibreather at frequency $\omega$. However, past $t = 0$, $\delta\theta_P$ begins to grow again, causing the field to diverge from this temporary quasibreather partner. This story repeats by choosing the next physical quasibreather to expand around at a nearby frequency $\omega + d\omega$, such that the attractive region of this new quasibreather has some overlap with the repulsive region of the previous quasibreather at $\omega$ (see figure \ref{fig:AE}). The term ``asymptotic attractor'' is chosen because of its likeness to the concept of asymptotic series, in which increasing the order of an expansion increases its precision until, at some point, it begins to diverge.

\subsection{\label{sec:EnergeticInstability}Energetic instability}

The physical quasibreather background around which we expand the perturbation $\delta\theta$ is one among a continuum of quasibreathers, parametrized by their fundamental frequency $\omega$. Thus, when we talk about a perturbation $\delta\theta$, we introduce the notation $\delta\theta^\omega$ in order to talk about ``the perturbation relative to (the deformed physical quasibreather of frequency) $\omega$,'' where we may omit the parenthetical when it is unambiguous to do so.

In the previous section, we introduced the concept of asymptotic attraction, in which an oscillon may be viewed as approaching a physical quasibreather for a finite period of time. For each quasibreather, there is an epoch of attraction, during which the particular solution $\delta\theta_P$ is shrinking towards zero, and an epoch of repulsion, during which $\delta\theta_P$ is growing away from zero. Neighboring quasibreathers at $\omega$ and $\omega + d\omega$ have particular solutions that cross zero at different absolute times $t = t_\omega$ and $t = t_{\omega + d\omega}$ respectively; whether $t_\omega < t_{\omega + d\omega}$ determines whether $\omega + d\omega$ is attractive for some time after $\omega$ becomes repulsive. Because the particular solution $\delta\theta_P$ encodes the energy flow out of the oscillon, the relative timing of the zero crossings of $\delta\theta_P^\omega$ and $\delta\theta_P^{\omega + d\omega}$ may be viewed as a reflection of energy conservation, defining an arrow of time. That is, the oscillon at $\omega + d\omega$ is energetically accessible from $\omega$ if $t_\omega < t_{\omega + d\omega}$. 

This time ordering implies the existence of a relative energy function, whose local monotonicity encodes whether $\omega + d\omega$ is accessible from $\omega$. In other words, the physical quasibreather at $\omega + d\omega$ is energetically accessible from $\omega$ if there is a time when $\delta\theta^\omega$ is a negative energy perturbation relative to $\omega$ and $\delta\theta^{\omega + d\omega}$ is a positive energy perturbation relative to $\omega + d\omega$. However, the energy of the total field configuration is ill-behaved because of the divergent radiative tails. Strictly speaking, because the quasibreather at $\omega$ has a different amplitude radiative tail than the quasibreather at $\omega + d\omega$, $\delta\theta$ cannot be a finite energy perturbation of both quasibreathers. However, the tails are decoupled and do not influence the dynamics of the oscillon bulk. Therefore, our measure of the perturbation energy must be agnostic to the radiation tails.

One might be inclined to count only the energy inside some finite box containing the oscillon bulk. However, such a measure still grows polynomially with the size of the box. One may also try to subtract the radiative tails by removing the $1/r$ (in $d = 3$) asymptotic, although again, this depends on an explicit cutoff between the bulk and the tails. Our framework provides a natural resolution to this ambiguity. Specifically, the orthogonal deformation $\theta_\te{OD}$ provides a measure of the radiative tail of $\theta_\te{PQB}$ valid everywhere. It is the energy associated with this orthogonal deformation that we subtract, leading to our definition of the bound energy in the PQB
\begin{align}
\notag
    E_B&\equiv \lim_{r\to\infty}\left[\int_0^r\diff V\p{\f12\dot\theta_\te{PQB}^2 + \f12(\nabla\theta_\te{PQB})^2 + V(\theta_\te{PQB})}\right.\\&\left.-\int_0^r\diff V\p{\f12\dot\theta_\te{OD}^2 + \f12(\nabla\theta_\te{OD})^2 + V(\theta_\te{OD})}\right]\,.
\end{align}
Note, because $\theta_\te{OD}$ and $\theta_\te{PQB}$ are out of phase, this difference will oscillate around an average value, reflecting the uncertainty principle. This definition has the virtue of converging to the deformed physical quasibreather energy when the oscillon is infinitely long-lived, i.e. when all harmonics are confined.

Having provided an unambiguous measure of the bound energy of the physical quasibreather, we may now address the question of when the perturbation $\delta\theta$ may flow between nearby quasibreathers. Because the frequency of the quasibreather, $\omega$, is a decreasing function of binding energy, under the assumption that binding energy decreases as total energy decreases, a family of physical quasibreathers is energetically stable where
\begin{align}
    \td{E_B}{\omega}{} < 0\,.
\end{align}

Inside a region of asymptotic attraction, the radiation power $P$ of the physical quasibreathers is a good approximation for the radiation power of the oscillon. This leads to a standard approximation for the oscillon lifetime, under the assumption that the perturbation $\delta\theta$ may completely relax (the adiabatic assumption)
\begin{align}
    T\approx \int\f{\diff E_B}{P(E_B)}\,.
\end{align}
As the oscillon becomes increasingly long-lived, and thus approaches the physical quasibreather, this prediction becomes increasingly precise.

\section{\label{sec:NM} Quasibreather numerical methods}
In the previous section, we arrive at the physical quasibreather as the main object of study which may be used to derive the properties of oscillons in a physical potential. In this section, we develop the numerical tools which enable the efficient calculations of physical quasibreathers and their orthogonal deformations.
\subsection{\label{sec:LR} Linear radiation}
Let us begin by supposing that $S_n$ and $C_n$ are known for $n< n_\te{pert}$, and that the remaining $S_n$ and $C_n$ are perturbatively small everywhere, so that they obey linear equations. Define the perturbation vector and the deformation vector respectively
\begin{align}
    \vec C = r^{(d - 1)/2}\mat{c}{C_{n_\te{pert}}\\ S_{n_\te{pert} + 1}\\\vdots}\,,\hspace{0.5cm}\vec s = r^{(d - 1)/2}\mat{c}{s_{n_\te{pert}}\\ c_{n_\te{pert} + 1}\\\vdots}\,,
\end{align}
the diagonal matrix of frequencies
\begin{align}
    \mathbf{\Omega}&=\mat{ccc}{n_\te{pert}\omega&\\&(n_{\te{pert}} + 1)\omega\\&&\ddots}\,,
\end{align}
the source vector $\vec {\mathcal J}(S_1,C_2,\dots)$ and the mass matrices $\mathbf{V}_{\vec C}(S_1,C_2,\dots),\mathbf{V}_{\vec s}(S_1,C_2,\dots)$, which are functions of the non-perturbative harmonics. Finally, we define the Sommerfeld operator $\mathbf{S}$, which together with the Dirichlet-Neumann 1D flat Laplacian acting on each diagonal block $\bm{\nabla}^2$ \footnote{In $d\neq1,3$ one must remember to add $(d-1)(d-3)/(4 r^2)$ to account for the change of variables.}, contains the Sommerfeld radiation condition (e), provided in appendix \ref{sec:PerturbativeHarmonicFormulae}. In this notation, the equations of motion for the perturbation and the deformation can be written as a sparse linear system
\begin{align}
\label{eq:LinearMatrix}
    \mat{c}{\vec {\mathcal J}_{\vec C}\\\vec{\mathcal J}_{\vec s}}&=\mat{cc}{-\mathbf{\Omega}^2 - \bm{\nabla}^2 + \mathbf{V}_{\vec C}&\mathbf{S}\\-\mathbf{S}&-\mathbf{\Omega}^2 - \bm{\nabla}^2 + \mathbf{V}_{\vec s}}\mat{c}{\vec C\\\vec s}\,.
\end{align}
This form, in which $\vec C$ and $\vec s$ only couple through the boundary term $\mathbf{S}$, is guaranteed because $\vec C$ on its own solves the equations of motion, and hence any backreaction from a perturbation $\vec s$ must come at second order. The explicit forms of $\mathbf{S}$, $\mathbf{V}_{\vec C}$, $\mathbf{V}_{\vec s}$ and ${\mathcal J}_{\vec C}$, ${\mathcal J}_{\vec s}$ are provided for several cases of interest in the appendix \ref{sec:PerturbativeHarmonicFormulae}. Note, ${\mathcal J}_{\vec s}$ is proportional to the orthogonal deformation of the non-perturbative modes, and is therefore zero when all radiative modes are perturbative.

The fact that we may write the equations for the perturbative modes as a well-determined system of equations is a reflection of the fact that the radiative boundary conditions and regularity conditions completely (and uniquely, for the linear modes) specify the physical quasibreather.

\subsection{\label{sec:NLR} Nonlinear harmonics}
The perturbative method in the previous section amounts to solving a sparse linear system, a process that is computationally efficient. Thus, given the knowledge of the non-perturbative background harmonics, we can compute the contribution of arbitrarily many additional harmonics at almost no computational cost.

Now we must do the dirty work of computing the nonlinear harmonics. Computing $n_\te{pert} - 1$ non-perturbative harmonics will amount to shooting a point particle in $n_\te{pert} - 1$ dimensions, and tuning its initial condition so that it lands on the saddle-top at the origin.

The physical quasibreather only feels the orthogonal deformation at second order, and therefore we may use the following procedure to compute the deformed PQB to leading order:
\begin{enumerate}
    \item Choose $S_n(0)$ and $C_n(0)$ from the $n_\te{pert} - 1$ dimensional space of initial conditions.
    \item Shoot the harmonics $S_n, C_n$ from $r = 0$ to some large finite radius $r_\te{out}$ by evolving the mode equations (a).
    \item Use the radiative boundary conditions (e) to convert the $S_n$ to $c_n$ and $C_n$ to $s_n$ at $r = r_\te{out}$.
    \item Shoot the $c_n,s_n$ back to the origin in the background resulting from step 2.
    \item Check for regularity at the origin for the $c_n$ and $s_n$, and regularity at $r_\te{out}$ for the bound harmonics. If not regular, adjust $S_n(0)$ and $C_n(0)$ and repeat from step 2. Since regularity is equivalent to minimizing the first derivative, this can be implemented by a variant of a binary search procedure, e.g. golden section search.
    \item Compute any number of perturbative harmonics using the procedure from the previous section in the background of the non-perturbative harmonics.
    \item To account for linear backreaction of the perturbative harmonics, re-shoot the non-perturbative harmonics in the background of the perturbative harmonics. This last step is repeated until convergence.
\end{enumerate}
In practice, it is helpful to break down the $n_\te{pert} - 1$ dimensional search into $n_\te{pert} - 1$ linear searches that are performed hierarchically.
The process of nonlinear shooting is sped up by precomputing the potential functions and using table-lookup. This kind of optimization is especially important when dealing with periodic potentials where repeatedly computing Bessel functions is costly.

\begin{figure}[t]
    \centering
    \includegraphics[width = \columnwidth]{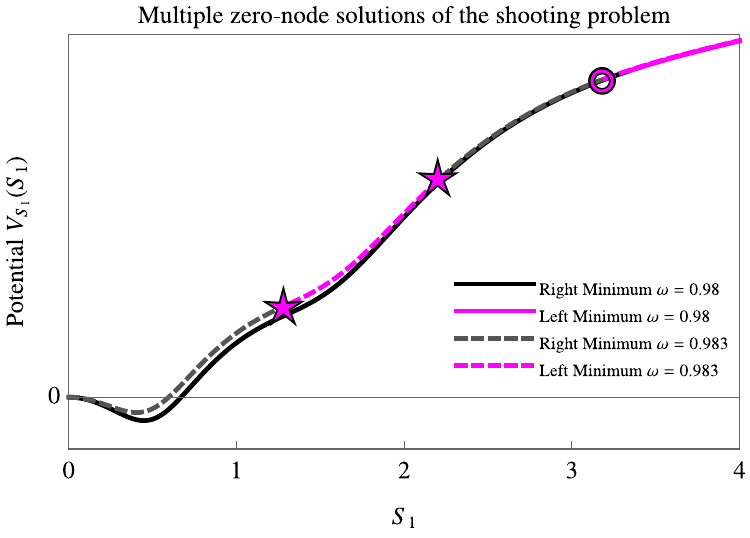}
    \caption{The emergence of two new zero-node solutions in the potential defined by Fourier coefficients $\vec{\mathcal V} = \{1, 0.5, -1, 0.5\}$ at large oscillon frequencies. The plot shows the effective potential $V_{S_1}(S_1)$ as a function of $S_1$ for positive values of $S_1$; since the potential is parity-symmetric, the $S_1<0$ region is the mirror opposite with respect to the $S_1=0$ axis. We have adjusted the vertical axis to better illustrate the qualitative features of the potential. Different regions are colored according to the sign of $S_1(\infty)$ when launched from that location. A shooting solution is represented by a point on the boundary between a black and magenta region. Whereas initially there was only one zero-node shooting solution (marked by the circle), the new potential adds two more zero-node solutions, marked by the stars. Intuitively, the higher the starting point, the further the particle will travel, causing successive solutions to have an increasing number of nodes. However, the combination of $2/r$ friction and nonlinearities in the potential breaks this intuition. Depending on the potential's convexity at the initial point, the oscillon may lose a widely variable amount of energy to friction. Therefore, it is at these regions of varying curvature that we expect these new solutions to emerge.}
    \label{fig:branching}
\end{figure}

\subsection{Branching of the fundamental mode}

In section \ref{sec:physicalQuasibreatherMinTech}, we reduce the problem of finding the radial profile of the oscillon to a classical-mechanical `shooting' problem. In its simplest case of one non-perturbative harmonic $S_1$, the problem further simplifies to the rolling of a massive ball in a double-welled potential $V_{S_1}(S_1)$ in the presence of $2/r$ friction. A shooting solution is one which starts at rest at an initial displacement $S_1(0)$ and ends at $S_1 = 0$ at $r=\infty$.

In linear equations, such as the radial hydrogen atom problem, there is exactly one solution for each integer number of nodes (i.e. zero-crossings) of the radial profile $S_1$. In our PQB mode equations, strong nonlinearities break this intuition, as depicted in figure \ref{fig:branching}. Specifically, a small change in the oscillon frequency $\omega$ can change the number of solutions with zero nodes by an increment of two, introducing new branches of oscillon solutions (or eliminating them) when the potential possesses non-trivial convexity. If this new branch consists of quasibreathers with lower binding energy, then the original branch may jump to the low-energy branch after the original branch experiences energetic death. In the reverse scenario, oscillons may form on the high-energy branch but the low-energy branch is energetically forbidden from reaching the high-energy branch.

In the oscillons that we study in e.g. figure \ref{fig:landscape1}, many of the longer-lived potentials contain a high-energy branch of very large, low-amplitude oscillons which only exists in a small range of frequencies close to $m$. One such example is shown in figure \ref{fig:branching}. All the examples studied in section \ref{sec:prescription} are the result of purposefully introducing these branches at a specific frequency $\omega \approx m_f$.

\section{\label{sec:FloquetAnalysis} Floquet analysis}

In appendix \ref{sec:ThePhysicalQuasibreatherFormalism} we introduced the notion of asymptotic attraction to describe physical oscillons as perturbations of PQBs. From this expansion, we have reduced the problem of oscillon stability to the study of the linear stability of equation \eqref{eq:perturbationHomEQ}. Standard Floquet theory tells us that the result of this analysis can have two outcomes: the equation is linearly unstable, or it possesses oscillatory states exclusively (modulo boundary effects). In other words, the existence of a stable decaying mode implies the existence of a growing mode, and stability must emerge at higher order in perturbation theory, if at all. Here, we address the linear stability of perturbations $\delta\theta$, and later argue that nonlinear terms stabilize linearly oscillatory modes.

\subsection{Linear stability analysis}

Let us begin by reproducing equation \eqref{eq:perturbationHomEQ} for ease of reference: the linearized homogeneous equation for the perturbation $\delta\theta$ in the background of $\theta_\te{PQB}$ is,
\begin{align}
\label{eq:homogeneous}
    0&=\delta\ddot\theta - \nabla^2\delta\theta + V''(\theta_\te{PQB})\delta\theta\,.
\end{align}
Recall that $\theta_\te{PQB}$ is a periodic solution of the equations of motion, and therefore can induce parametric resonances.
Substituting in the form of the quasibreather \eqref{eq:QB}, we find
\begin{align}
\label{eq:Pert}
    &V''(\theta_\te{PQB})=\\\notag& \sum_{m\in\N_\te{even}}V_m''(S,C)\cos m\omega t+\sum_{m\in\N_\te{odd}}V_m''(S,C)\sin m\omega t\,,
\end{align}
which, under parity symmetry of $V$, further simplifies to
\begin{align}
\label{eq:PertParity}
     V''(\theta_\te{PQB})&=\sum_{m\in\N_\te{even}}V_m''(S)\cos n\omega t\,,
\end{align}
where $V_m''$ is defined by \eqref{eq:Pert}.
We will leave specific formulae for $V_m''$ to appendix \ref{sec:LinStabEquations}.

Since (\ref{eq:perturbationHomEQ}) is linear, we may Fourier transform $t\to \Omega$, and decompose $\delta\theta$ in spherical harmonics. Because the quasibreather background is periodic, it induces couplings between frequencies separated by integer multiples of the fundamental frequency $\omega$. Therefore, let us restrict our analysis to the values of the Fourier transform $\delta\theta(\Omega, r)$ at the discrete tower of harmonics defined as $\Omega_n\equiv \Omega_0 + n\omega$, $n\in\Z$, where the base frequency $\Omega_0$ can be assumed to lie in the interval $(0, \omega)$. Therefore, the Fourier components on this tower, denoted $\delta\theta_n(\Omega_0,r) \equiv \delta\theta(\Omega_n, r)$, will respect a matrix-differential equation:
\begin{widetext}
\begin{align}
\label{eq:fullLinearPert}
\notag
    0&=-(\Omega_0 + n\omega)^2\delta\theta_n - \delta\theta_n'' - \f{d - 1}{2}\delta\theta_n' + \f{\ell(\ell + d - 2)}{r^2} \delta\theta_n + V_0''(S,C)\delta\theta_n+\\&+ \f12\sum_{m\in\N_\te{even}^{>0}}V_m''(S,C)(\delta\theta_{n + m} + \delta\theta_{n - m}) +\f1{2\I}\sum_{m\in\N_\te{odd}}V_m''(S,C)(\delta\theta_{n + m} - \delta\theta_{n - m})\,,
\end{align}
\end{widetext}
where $\ell$ is the angular momentum number. 
This is apparently a quadratic eigenvalue problem in the fundamental frequency $\Omega_0$ \cite{tisseur2001quadratic}, although as we will see, it becomes an irrational eigenvalue problem upon imposing transparent boundary conditions \cite{sitnikov2003siegert,ostrovsky2005scattering}.
The eigenvalue solutions $\Omega_0$ characterize the stability or instability of the system: real eigenvalues corresponding to oscillatory motion; if solutions pick up an imaginary part, the mode will be exponentially growing (if $\Im(\Omega_0) < 0$) or exponentially decaying (if $\Im(\Omega_0) > 0$). In the absence of transparent boundary conditions, the solutions come in pairs of complex conjugates; in this closed-box scenario, the existence of a stable (i.e. decaying) mode implies the existence of an unstable mode.

\textit{An instructive example}\hspace{0.3cm}
We may gain some insight into the eigenvalues $\Omega_0$ by studying the simpler case of perturbations inside a box for a potential with parity. The matrix differential equation simplifies such that only the sum over even terms in \eqref{eq:fullLinearPert} survives.  At leading order we only include the first harmonic's $n = \pm1$ terms; the reason is that large-$n$ harmonics both decouple from the fundamental and become unbound. This allows us to keep only the $V_0''$ and $V_2''$ terms in the equation. Moreover, we make the assumption that $\Omega_0$ is small compared to $\omega$, representing solutions to the perturbation equations with a separation between the fast and slow timescales; this is relevant when we focus on the boundary between periodicity and instability, where $\Omega_0$ will be small in magnitude. This assumption will be supported by the result of the analysis.

The result is the following $2\times2$ matrix-differential equation
\begin{align}
    0&=\ps{2\omega\Omega_0\bm{\sigma}_z + \bm{\sigma}_x V_2'' + \mathbf{I}(\omega^2 - \nabla^2 + V_0'')} \delta \vec\theta\,,
\end{align}
where we have suppressed the argument of $V_0'',V_2''$ for brevity and $\bm{\sigma}_i$ are the Pauli matrices with entries of magnitude 1. 
We may phrase this as a typical eigenvalue problem in $\Omega_0$ by multiplying through with $\bm{\sigma}_z/2\omega$, leading to
\begin{align}
\label{eq:simplifiedEigensystem}
    0 &= \det\p{\mathbf{H} + \mathbf{A} - \Omega_0\mathbf{I}}\,,
\end{align}
where $\mathbf{H}$ and $\mathbf{A}$ are Hermitian and anti-Hermitian matrices defined by
\begin{align}
    \mathbf{H}&=\f{1}{2\omega}\bm{\sigma}_z(\omega^2 -\nabla^2 + V_0'')\,,\hspace{0.5cm}\mathbf{A}= \f{\I}{2\omega}\bm{\sigma}_y V_2''\,.
\end{align}

\begin{subfigures}
\begin{figure}[t]
    \centering
    \includegraphics[width = \columnwidth]{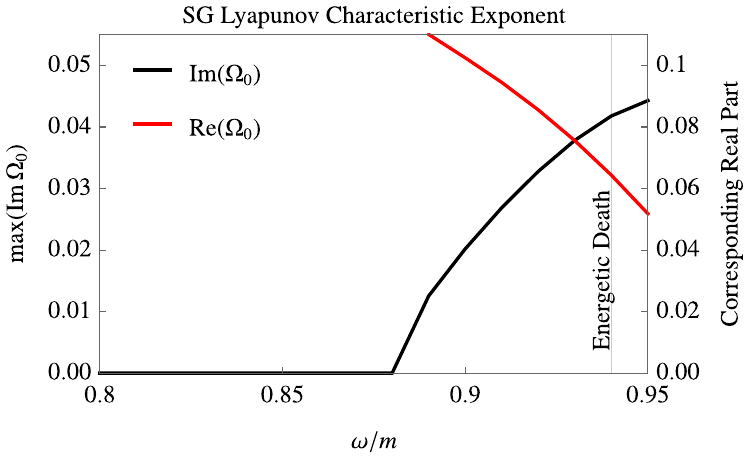}
    \caption{The Lyapunov characteristic exponent (the eigenvalue $\Omega_0$ of \eqref{eq:fullLinearPert} with maximum imaginary part) for the sine-Gordon deformed physical quasibreather (with an error of $\pm 0.005$). The perturbation $\delta\theta$ becomes linearly unstable at $\omega\approx 0.88$. The nearest asymptotically attractive quasibreather is always finitely far away from the oscillon. When $\omega > 0.88$, the linearly unstable mode is therefore \emph{always} excited, leading to growing quasiperiodic oscillations on top of the deformed quasibreather background (see figure \ref{fig:SGEvo}). Note, throughout this band of linear instability, the mass energy $\int\diff V \f14 m^2 S_1^2$ is monotonically decreasing, in contradiction with \cite{amin2010flat}. On the plot, we denote the energetic death at $\omega\approx 0.94$, where the oscillon is forced off the quasibreather trajectory by energy conservation.}
    \label{fig:SGLinStab}
\end{figure}
\begin{figure}[t]
    \centering
    \includegraphics[width = \columnwidth]{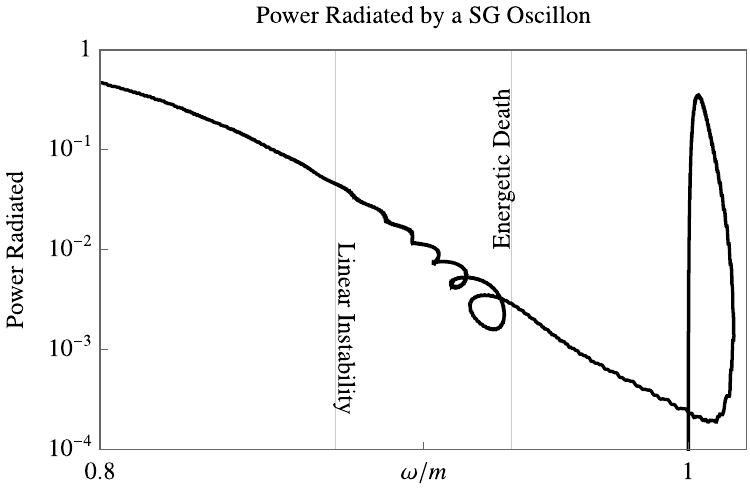}
    \caption{The power radiated by a simulated sine-Gordon oscillon versus the central fundamental frequency. On this plot, we've indicated the onset of linear instability $\omega\approx 0.88$ calculated using our eigenvalue code described in appendix \ref{sec:FloquetAnalysis}, and the instance of energetic death $\omega\approx 0.94$ described in appendix \ref{sec:EnergeticInstability}. This figure represents the a consequence of linear instability: growing quasiperiodic oscillations. The specific magnitude of this effect depends on initial conditions and environmental perturbations (see figure \ref{fig:SG} for an example where oscillations are suppressed). Whether or not the unstable mode can become large enough to destroy the oscillon, the perturbation itself has a radiation component, which may significantly modify the lifetime. In this particular case, the unstable mode's frequency $\omega \pm \te{Re}\Omega_0$ approaches the oscillon frequency $\omega$ towards the end of life, leading to growing beats (see figure \ref{fig:SGLinStab}). The loop of death at the end of the evolution occurs because the central oscillon rapidly becomes a mix of first and third harmonic, causing the central frequency to be larger than 1.}
    \label{fig:SGEvo}
\end{figure}
\end{subfigures}

In other words, the $\Omega_0$ eigenvalues are the roots of the characteristic polynomial with all-real coefficients defined by the matrix with all-real entries $\mathbf{H} + \mathbf{A}$. Consider the case $\mathbf{A}=0$; the eigenvalues $\Omega_0$ are the eigenvalues of $\mathbf{H}$, which is composed of two mirrored copies of the real spectrum of the single-block operator $\f{1}{2\omega}(\omega^2 -\nabla^2 + V_0'')$. The addition of $\mathbf{A}$ only introduces couplings between these two sectors; since it is also antisymmetric, these couplings are equal and opposite in sign. If we start from a spectrum of $\mathbf{H}$ with no overlap between its two sectors, the addition of $\mathbf{A}$ will bring the two mirrored \lq ground states\rq\ together from $\Omega_0=\pm E_\te{ground}$ to the value of $\Omega_0=0$. From the perspective of the characteristic polynomial, this corresponds to the two roots becoming degenerate before turning imaginary. In other words, complex eigenvalues must appear by first passing through an inter-block degeneracy. Therefore, the meeting of the two ground states defines the boundary between periodicity (i.e. an all-real spectrum) and instability (i.e. complex spectrum). If the spectrum of $\mathbf{H}$ is bounded below by 0, then the meeting of the two states will produce purely imaginary eigenvalues. This result should be compared to \cite{amin2010flat}. In general, the symmetries of \eqref{eq:simplifiedEigensystem} that led to this result are only approximate, and therefore we should expect the first nearly-stable eigenvalue to be close to zero in general.
\\\\
\indent
However, because the oscillon lives in an open box, we must ensure that \eqref{eq:fullLinearPert} is endowed with transparent boundary conditions. Such radiative boundary conditions depend on the momentum of the outgoing mode  $\sqrt{(\Omega_0 + n\omega)^2 - 1}$. Eigenvalue problems with radiative boundary conditions have been studied in the non-relativistic limit in \cite{sitnikov2003siegert}. Crucially, the calculations of \cite{sitnikov2003siegert} depend on the existence of a uniformizing variable $u(\Omega_0)$ in which the outgoing momentum becomes a rational function of $u$. As far as we are aware, no such uniformization procedure is known for the relativistic case with two channels, or more generally for any case with more than two channels.

Using series approximations and a uniformizing variable $u(\Omega_0)$, we show in appendix \ref{sec:LinStabEquations} that it is possible to express the boundary condition as a polynomial for $\abs{\Omega_0} < 1/2$ for $\omega > 1/2$ in the case of parity or $\omega > 3/4$ without parity. Using standard linearization techniques, we may reduce this polynomial eigenvalue problem to a generalized eigenvalue problem, for which numerical software is plentiful. This is the method applied to the stability analysis in Figure \ref{fig:SGLinStab}.

\begin{figure}
    \centering
    \includegraphics[width=\columnwidth]{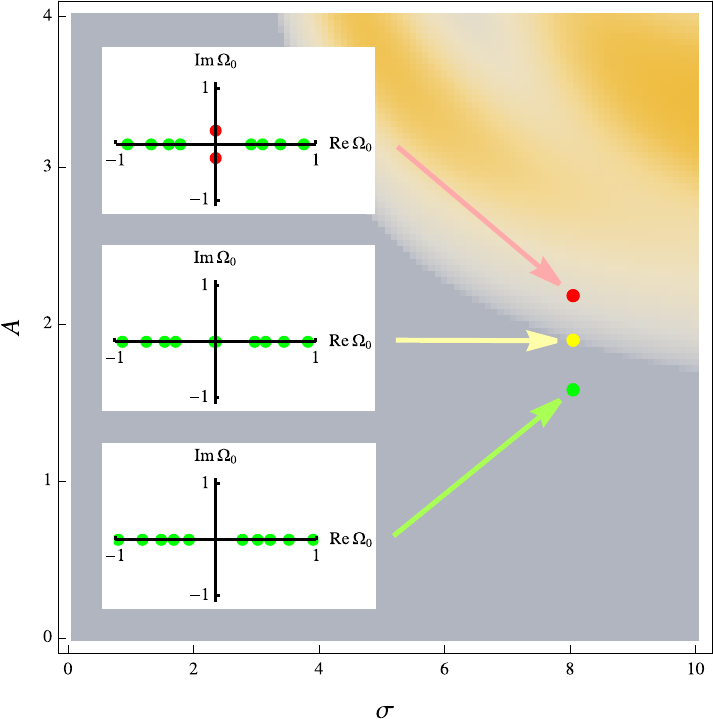}
    \caption{A visualization of how linear instability emerges in the simplified model of appendix \ref{sec:FloquetAnalysis}. The boundary of stability is described by eigenvalues meeting at zero. The plot describes the solutions to the eigenvalue equation \eqref{eq:simplifiedEigensystem} in the case of a simple Gaussian background, in which the fundamental oscillon mode is taken to be $S_1(r)=A\exp{-r^2/2\sigma^2}$. The plot background describes stability as a function of the two Gaussian parameters, the oscillon amplitude $A$ and width $\sigma$; for oscillons of sufficient width and amplitude, there are eigenvalues $\Omega_0$ with negative imaginary part, and thus the oscillon is unstable. We show the eigenvalues nearest to zero for three points in this parameter space: stable (green), borderline unstable (yellow), and unstable (red). The real eigenvalues closest to the origin become degenerate at zero on the boundary of stability; they further split into purely imaginary conjugates in the instability region.}
    \label{fig:InstabilityBoundary}
\end{figure}

\subsection{Nonlinear stabilization}
\begin{figure}
    \centering
    \includegraphics[width = \columnwidth]{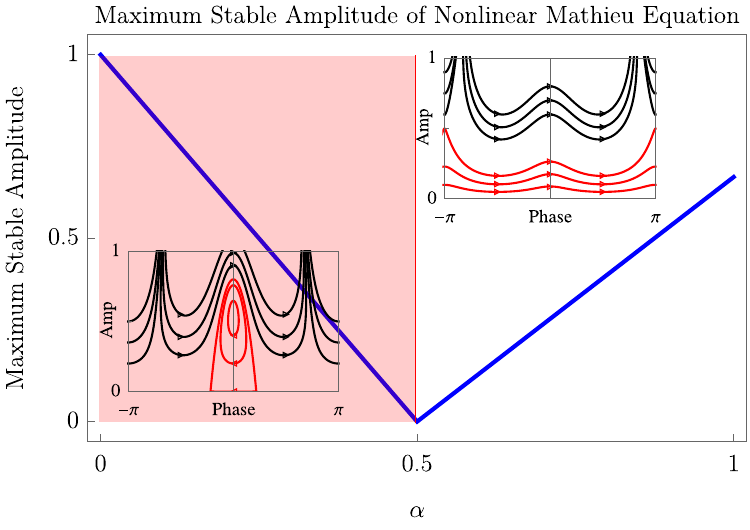}
    \caption{Here we plot the maximum stable amplitude of $y$ in the nonlinear Mathieu equation \eqref{eq:MathieuNL} for small $\epsilon$, and we've indicated the instability band of the linear Mathieu equation in red. Outside the red region, the nonlinear oscillations are centered on $y = 0$, representing that the oscillations stay bounded independent of phase. However, for $\abs{\alpha} < 0.5$, only oscillations of a particular phase remain bounded, indicating that $y = 0$ has become hyperbolic (see left inset). Inset in the plot are two examples of the slow oscillation trajectories. For $\abs{\alpha} < 0.5$, the red stable trajectories have amplitude larger than 0 and are restricted to a finite interval of phase. This generally nonlinear phenomenon represents a special region of stability within the otherwise unstable phase of the Mathieu parameter space. For $\abs{\alpha} > 0.5$, the red stable oscillations are restricted to a finite amplitude, but are allowed to have any phase. In both cases, large enough amplitude perturbations grow without bound, represented by the black trajectories.}
    \label{fig:NonlinearMathieu}
\end{figure}
In the previous section, we laid out our numerical method for computing the linear stability of the homogeneous perturbation $\delta\theta$ in the background of the deformed physical quasibreather. Modulo technicalities at the boundary, we found that all linear perturbations either are oscillatory, or come in pairs of exponentially growing and decaying modes. The result is that stability, understood to mean that all homogeneous perturbations shrink, cannot be fully explained at the level of linear Floquet analysis.

Thus, stability must originate at higher order in perturbation theory, if it exists at all. In this section, we identify radiation as the mechanism of stabilization accessible to small oscillatory perturbations; specifically, modes which are periodic in the linear stability analysis will couple to radiative modes at higher orders, providing a channel for dissipation. Therefore, we conclude that a sufficient condition for full nonlinear stability is that all modes are oscillatory at the level of linear perturbation theory. Furthermore, we will see that linear instability does \emph{not} imply nonlinear instability.

We will explore the effect of adding a nonlinear term to a Floquet-type problem by studying the toy example of the Mathieu equation with a quadratic nonlinear term. To simplify our analysis, we begin by studying potentials with parity, so that the leading order oscillating contribution to $V''(\theta_\te{PQB})$ is proportional to $\cos2\omega t$. The leading nonlinear term is then proportional to $\sin\omega t$. Thus, we will study the nonlinear generalizations of the Mathieu equation of the form
\begin{align}
    \label{eq:MathieuNL}
    0&=\ddot y + (1 +\epsilon( \alpha + \cos2 t +  y\sin t)) y\,.
\end{align}
In this toy problem, $y$ represents the perturbation $\delta\theta$ to a physical quasibreather whose potential conserves parity. The fact that the linear term is proportional to $\cos2t$ and the quadratic term $y^2$ is proportional to $\sin t$ is a consequence of the symmetry of the potential, which guarantees that polynomials in $y$ of certain parity have the corresponding oscillatory terms.

A standard two-timing analysis, along the lines of \cite{jordan1973two}, with $\epsilon\ll 1$ demonstrates that the Mathieu instability bifurcation at $\abs{\alpha}=1/2$ is unchanged by the nonlinearity around $y = 0$. However, one difference is the appearance of regions of stability inside the linearly unstable region $\abs{\alpha}<1/2$, although large enough $y$ always implies instability, regardless of $\alpha$ \footnote{Indeed, this example is illustrative for much of the behavior we observe in numerical simulations of oscillons. Large quasiperiodic oscillations, arising from a linear instability, do not blow up like the linear analysis would suggest, but rather persist because of nonlinear regions of stability.}. When $y$ is the smallest scale in the problem, we recover the usual Mathieu equation behavior (see figure \ref{fig:NonlinearMathieu}). In summary, linear periodicity is unchanged for small enough $y$, although linearly unstable modes may become periodic.

Thus, we should expect that the oscillatory modes of the linear equation \eqref{eq:fullLinearPert} remain oscillatory upon introduction of a nonlinear term as long as they are of small enough amplitude. Moreover, the nonlinear terms may convert an otherwise linearly unstable mode into an oscillatory one. Further, the nonlinear interactions of linearly oscillatory modes will necessarily produce radiation, carrying away energy, causing their amplitude to shrink. Thus, sufficiently small linearly oscillatory modes are stabilized by radiation.

\subsection{Angular perturbations}

\begin{figure}
    \centering
    \includegraphics[width=\columnwidth]{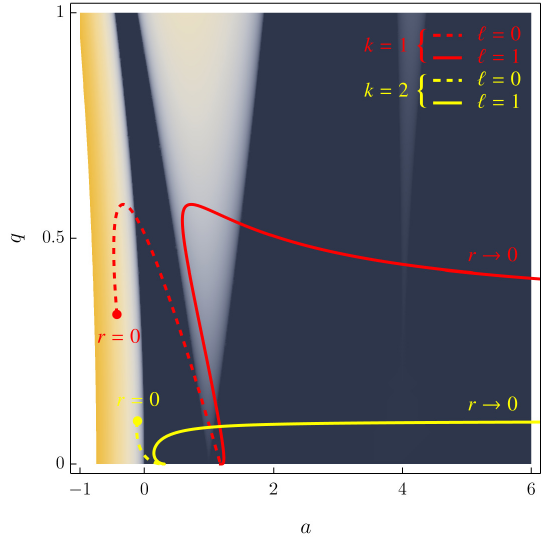}
    \caption{Effective Mathieu equation parameters $0 = \ddot y + (a - 2 q\cos2 k \omega t) y$ for integer $k$, where we associate a pair $(a_r,q_r)$ to each radius $r$ of the sine-Gordon quasibreather background \eqref{eq:fullLinearPert} for $\omega = 0.95$, ignoring the gradient term. This picture is meant to guide our intuition of the Mathieu equation into the less-familar Floquet problem \eqref{eq:fullLinearPert}. Intuitively, a mode can be understood as more unstable if more of its volume lies in the Mathieu instability bands. This plot, although not quantitatively precise, provides intuition for why the lowest angular momentum states are more susceptible to instabilities, since they have the most overlap with the dominant instability bands.}
    \label{fig:Mathieu}
\end{figure}

In appendix \ref{sec:LinStabEquations}, we develop a calculation scheme to solve for the perturbation $\delta\theta$ as a function of $t$ and $r$. In order to obtain the perturbation equations for $\delta\theta$, we performed a spherical harmonic decomposition, resulting in a set of decoupled equations for each mode of angular momentum number $\ell$. These equations differ by the coefficient of the angular momentum effective potential
\begin{align}
    V_\te{angular} &= \f{\ell(\ell + d - 2)}{r^2}\,.
\end{align}
Because this potential is positive, it acts as a repulsive centrifugal term, reducing the perturbation density at the origin. Hence, we expect that perturbations with more angular momentum are typically {\em more linearly stable}, since less of the perturbation lies inside the oscillon bulk, although for low angular momentum, the conclusion is case-dependent. An intuition for this comes from applying the stability phases of the standard Mathieu equation (see figure \ref{fig:Mathieu}).

A similar $1/r^2$ term appears in the effective potential for the perturbation upon removing the $(d-1)/r$ friction term through a change of variables $\delta\theta\to r^{(d-1)/2}\delta\theta$. This introduces the effective potential
\begin{align}
    V_\te{geometric} &= -\f{(d-1)(d-3)}{4r^2}\,.
\end{align}
This term differs from the angular momentum term in two important ways. First, it can be of either sign: for $d = 1,3$ it vanishes, for $d = 2$, it is repulsive, and for $d\geq 4$ it is attractive. Second, it also influences the quasibreather background itself, whereas the angular momentum terms only influence the non-spherical perturbations. Because this potential influences both the background \emph{and} the perturbation, its effect on stability depends on the specifics of the nonlinear potential.

\section{\label{sec:TechnicalFormulae} Technical formulae}
In this section, we provide a detailed description of the formulae and numerical techniques used to compute physical quasibreather properties.

\subsection{\label{sec:PerturbativeHarmonicFormulae}Perturbative harmonic formulae}
Once we have computed the oscillon's non-perturbative modes $S_{n<n_\te{pert}}, C_{n < n_\te{pert}}$ and their orthogonal deformations $c_{n<n_\te{pert}}, s_{n < n_\te{pert}}$, we may compute the perturbative modes $S_{n \geq n_\te{pert}}, C_{n \geq n_\te{pert}}$ and their orthogonal deformations $c_{n\geq n_\te{pert}}, s_{n \geq n_\te{pert}}$ using the procedure outlined in section \ref{sec:LR}. Here, we provide explicit formulae for the Sommerfeld matrix $\mathbf{S}$ and the block Laplacian operator $\bm{\nabla}^2$. 
Upon discretizing space such that $r = [0,\diff r,\dots,(N-1)\diff r]$, the matrix $\mathbf{S}$ is comprised of all zeros, except for the lower right entry in each diagonal block. To describe this object, we introduce the following notation. The matrix $\mathbf{S}$ has four indices: two upper indices labelling the block, and two lower indices labeling the location within that block. In this notation, the entries in the matrix $\mathbf{S}$ in \eqref{eq:LinearMatrix} may be written
\begin{align}
    \mathbf{S}_{ab}^{nm}&=(-1)^{n}\sqrt{(n\omega)^2 - 1}\delta_{nm}\delta_{aN}\delta_{bN}\,.
\end{align}
In the same notation, we may describe the Dirichlet-Neumann block Laplacian operator
\begin{align}
    [{\bm{\nabla}^2}]^{nm}&=\f{1}{\diff r^2}\mat{ccccc}{-2&1\\1\\&&\ddots&1\\&&1&-2&\diff r\\&&&1&-\diff r}\delta_{nm}\,.
\end{align}

\subsection{Potentials with parity}
As described in section \ref{sec:tuned?}, the Fourier basis is a natural basis to describe any scalar potential, since it is not plagued by the same radius-of-convergence issues of, say, the Taylor basis. Here we provide the harmonic factorization of a general scalar potential with parity \eqref{eq:PeriodicCoeff}. Taking the first derivative of \eqref{eq:PeriodicCoeff} with respect to $\theta$, we arrive at the following expression for the self-interaction terms in the non-linear wave equation
\begin{align}
    V'(\theta) = \sum_{m = 1}^\infty\f{{\mathcal V}_m}{m}\sin m\theta\,.
\end{align}
This expression is specific to the case of a potential with $2\pi$ periodicity. To accommodate potentials without periodicity, simply replace $\theta\to \theta/\theta_\te{max}$ where $\theta\in[-\pi\theta_\te{max},\pi\theta_\te{max}]$ and $\sum {\mathcal V}_m = \theta_\te{max}^2$. In order to keep the following expressions from getting any more unruly, we will present the explicit formulae for $2\pi$-periodic potentials, since the reader may easily convert these expressions to accommodate general periodicity.

By virtue of the Jacobi-Anger expansion \cite{dattoli1996theory}
\begin{align}
    e^{\I a\sin b} = \sum_{k = -\infty}^\infty J_k(a)e^{\I k b}\,,
\end{align}
we may compute the harmonic expansion of the potential, evaluated as a function of the PQB harmonics
\begin{align}
&V'\p{\sum_{n = 1}^\infty S_n\sin(n\omega t)}\\\notag& = \sum_{m = 1}^\infty\f{{\mathcal V}_m}{m}\sum_{\vec k\in \vec\Z}\ps{\p{\prod_{n = 1}^\infty J_{k_n}(m S_n)}\sin\p{\sum_{n = 1}^\infty nk_n\omega t}}\,,
\end{align}
where $\vec k = (k_1,k_2,\dots)$. One may write this more compactly in terms of generalized Bessel functions \cite{dattoli1996theory}. From this formula, we obtain the expressions for $V_{S_n}$ in \eqref{eq:ModeEQ}
\begin{align}
V'(\theta) \equiv \sum_{n = 1}^\infty V_{S_n}'(S_1,\dots,)\sin(n\omega t)\,.
\end{align}
In general, we may evaluate the full non-perturbative formulae for the mode-potential derivatives $V_{S_N}'$ and $V_{c_N}'$,
\begin{widetext}
\begin{equation}
\begin{aligned}
V_{S_N}' &= \sum_{m = 1}^\infty \f{{\mathcal V}_m}{m}\sum_{k_1^s,\dots, k_1^c,\dots}\p{\prod_{n = 1}^\infty J_{k_n^s}(m S_n) J_{k_n^c}(m c_n)}\cos\p{\sum_{n = 1}^\infty k_n^c \pi/2}\ps{\delta\p{N - \sum_{n = 1}^\infty  n(k_n^s  + k_n^c) } - \delta\p{-N - \sum_{n = 1}^\infty  n(k_n^s  + k_n^c) } }\,,\\
V_{c_N}' &= \sum_{m = 1}^\infty \f{{\mathcal V}_m}{m}\sum_{k_1^s,\dots, k_1^c,\dots}\p{\prod_{n = 1}^\infty J_{k_n^s}(m S_n) J_{k_n^c}(m c_n)}\sin\p{\sum_{n = 1}^\infty k_n^c \pi/2}\ps{\delta\p{N - \sum_{n = 1}^\infty  n(k_n^s  + k_n^c) } + \delta\p{-N - \sum_{n = 1}^\infty  n(k_n^s  + k_n^c) }}\,.
\end{aligned}
\end{equation}
\end{widetext}
Note, the $\delta$s in this equation are Kronecker $\delta$s, but we use a parenthetical argument to keep the expression readable. From these expressions, we may derive useful formulae for important cases of interest. Here, we present two examples for illustration, and because the reader may find them particularly useful in generating oscillon profiles of their own. First, in the case that the fundamental mode $S_1$ dominates and all other modes are perturbative, we have the following source term
\begin{align}
    \vec {\mathcal J}_{\vec C} = r^{(d - 1)/2}\sum_{m = 1}^\infty2 \f{{\mathcal V}_m}{m} \mat{c}{ J_{3}(m S_1) \\ J_{5}(m S_1) \\\vdots}\,,
\end{align}
with $\vec{\mathcal J}_{\vec s} = 0$ and the following mass matrices
\begin{widetext}
\begin{equation}
\label{eq:S3Matrix}
\begin{aligned}
    \mathbf{V}_{\vec C} &=\sum_{m = 1}^\infty {\mathcal V}_m \mat{ccc}{ \p{J_{3 - 3}(m S_1) - J_{3 + 3}(m S_1) }&\p{J_{3 - 5}(m S_1) - J_{3 + 5}(m S_1) }&\cdots\\
\p{J_{5 - 3}(m S_1) - J_{5 + 3}(m S_1) }& \p{J_{5 - 5}(m S_1) - J_{5 + 3}(m S_1) }&\\
\vdots& &\ddots}\,,\\
\mathbf{V}_{\vec s} &= \sum_{m = 1}^\infty {\mathcal V}_m \mat{ccc}{ \p{J_{3 - 3}(m S_1) + J_{3 + 3}(m S_1) }&\p{J_{3 - 5}(m S_1) + J_{3 + 5}(m S_1) }&\cdots\\
\p{J_{5 - 3}(m S_1) + J_{5 + 3}(m S_1) }& \p{J_{5 - 5}(m S_1) + J_{5 + 3}(m S_1) }&\\
\vdots& &\ddots}\,,
\end{aligned}
\end{equation}
\end{widetext}
to be inserted into equation \eqref{eq:LinearMatrix}. 
The case where $S_1$ and $S_3$ are non-perturbative and all other harmonics are perturbative everywhere has a similarly clean form
\begin{widetext}
\begin{equation}
\begin{aligned}
    \mathbf{V} &=\sum_{m = 1}^\infty V_m \sum_{k = -\infty}^\infty J_{k}(m S_3)\mat{ccc}{ \p{J_{5 - 5 - 3 k}(m S_1) \mp J_{5 + 5 - 3 k}(m S_1) }&\p{J_{5 - 7 - 3 k}(m S_1) \mp J_{5 + 7 - 3 k}(m S_1) }&\cdots\\
\p{J_{7 - 5 - 3 k}(m S_1) \mp J_{7 + 5 - 3 k}(m S_1) }& \p{J_{7 - 7 - 3 k}(m S_1) \mp J_{7 + 7 - 3 k}(m S_1) }&\\
\vdots& &\ddots}\,,\\
    \vec {\mathcal J}_{\vec C} &= r^{(d - 1)/2}\sum_{m = 1}^\infty \f{V_m}{m} \sum_{k = -\infty}^\infty J_{k}(m S_3)\mat{c}{ J_{5 - 3 k}(m S_1) - J_{-5 - 3k}(m S_1) \\  J_{7 - 3 k}(m S_1) - J_{-7 - 3k}(m S_1) \\\vdots}\,,\\
    \vec {\mathcal J}_{\vec s} &= r^{(d - 1)/2}c_3 \sum_{m = 1}^\infty V_m\sum_{k = -\infty}^\infty J_{k - 1}(m S_3)\mat{c}{ J_{5 - 3 k}(m S_1) + J_{-5 - 3 k}(m S_1) \\  J_{7 - 3 k}(m S_1) + J_{-7 - 3 k}(m S_1) \\\vdots}\,,
\end{aligned}
\end{equation}
\end{widetext}
with $-$ corresponding to $\mathbf{V}_{\vec C}$ and $+$ corresponding to $\mathbf{V}_{\vec s}$. The formulae when there are more non-perturbative harmonics follow the same pattern.
\subsection{\label{sec:LinStabEquations}Formulae for linear stability analysis}
Here we provide the mathematical details to accompany appendix \ref{sec:FloquetAnalysis}. We restrict ourselves to $\omega > 1/2$ in the case of parity and $\omega > 3/4$ otherwise. This restriction is to ensure that the following series approximation converges on the disc $\abs{\Omega_0} < 1/2$. If one is certain that the unstable eigenvalues occur in a smaller disc, the restrictions on $\omega$ may be weakened significantly, and indeed, this is often the case.

The outgoing boundary conditions depend on the momentum of the outgoing mode, which is an irrational function of $\Omega_0$. In order to convert the irrational eigenvalue problem \eqref{eq:fullLinearPert} into a polynomial eigenvalue problem that may be solved with standard techniques, we need to approximate the momentum $\sqrt{(\Omega_0 \pm n\omega)^2 - 1}$ by a polynomial. One's first intuition might be that the Taylor series of the momentum expanded about $\Omega_0 = 0$ would be a good approximation. This intuition is good for the higher harmonics, since $\Omega_0$ is often much smaller than $n\omega$. However, this series only converges inside the disc $\abs{\Omega_0} < 1-\omega$ for $n = 1$, which is not sufficient to compute the Lyapunov exponent of the linear perturbation $\delta\theta$. A more sophisticated approximation is necessary in order to capture the behavior of the momentum as a function of $\Omega_0$ on a disc that remains finite size as $\omega\to 1$.

To this end, we define 
\begin{align}
    x = \Omega_0^2 - (1 - \omega)^2\,,
\end{align}
so that 
\begin{align}
    \sqrt{(\Omega_0\pm\omega)^2 - 1} = \sqrt{x + 2\omega(\omega - 1\pm\Omega_0)}\,.
\end{align}
We then Taylor expand around $x = 0$, yielding the following series that converges on the disc of radius $1/2$ centered on $\Omega_0 = 0$ for $\omega > 1/2$
\begin{widetext}
\begin{equation}
\label{eq:expansions1}
    \begin{aligned}
    \sqrt{(\Omega_0 \pm \omega)^2 - 1}&=\sqrt{2\omega(\omega - 1\pm \Omega_0)}\ps{1-\sum_{j = 0}^\infty\mat{c}{2j\\j}\f{1}{(j + 1)2^{2j + 1}}\p{\f{(1 - \omega)^2-\Omega_0^2}{2\omega(\omega - 1\pm \Omega_0)}}^{j + 1}}\,,\\
    \sqrt{(\Omega_0 \pm n\omega)^2 - 1}&=\sqrt{(n\omega)^2 - 1}\ps{1 - \sum_{j = 0}^\infty\mat{c}{2j\\j}\f{1}{(j + 1)2^{2j + 1}}\p{-\f{(\Omega_0\pm n\omega)^2-(n\omega)^2}{(n\omega)^2 - 1}}^{j+1}}\,,
    \end{aligned}
\end{equation}
\end{widetext}
where the second equation is just the ordinary Taylor expansion centered on $\Omega_0 = 0$ for $n\geq 2$.
The factor $\sqrt{2\omega(\omega - 1\pm \Omega_0)}$ is not yet a polynomial. We utilize the technique of uniformization \cite{sitnikov2003siegert}, where we define the complex variable $u$ such that $\sqrt{2\omega(\omega - 1\pm \Omega_0)}$ becomes a polynomial in $u$,
\begin{align}
    \Omega_0 = \f{1-\omega}{2}\p{u^2 + u^{-2}}\,.
\end{align}
This definition turns \eqref{eq:expansions1} into rational functions of $u$, allowing us to rephrase \eqref{eq:fullLinearPert} as a polynomial eigenvalue problem. The zero eigenvalues now live at the four roots of the equation $u^4 = -1$, around which we perform a small eigenvalue search using the Krylov subspace methods implemented in \texttt{Matlab}. Once the $u$ eigenvalues and eigenvectors have been computed, we must convert back to check that they correspond to eigenvalues of the original irrational eigenvalue problem. In short, we have reduced the original irrational eigenvalue to a polynomial eigenvalue problem of degree at-least 4, depending on the degree of accuracy one wants to achieve.

Finally, we define the matrix $\mathbf{S}$ which encodes the non-derivative term in the Sommerfeld radiation condition, which can be written as a sum of the matrices $\hat{\mathbf{S}}_\lambda$, which are matrices of all zeros except for the lower right entry of the $\lambda$th diagonal block, which is 1. This entry corresponds to the outer boundary of the grid, with $\lambda$ ranging from $-L$ to $L$, and the upper left block corresponding to $\lambda = -L$, where $L$ is the chosen order of the Floquet expansion. In this notation, the non-derivative part of the Sommerfeld boundary conditions may be written
\begin{align}
\mathbf{S} &= \sum_{\lambda = -\infty}^\infty\sqrt{(\Omega_0 + \lambda\omega)^2 - 1}\hat{\mathbf{S}}_\lambda\approx \sum_{\lambda = -L}^L\sum_{i =0}^{2 M + 1} c_{\lambda,i} u^i\hat{\mathbf{S}}_\lambda\,,
\end{align}
where $M = \te{min}(2j_{\te{max},1} + 3,4 j_{\te{max},n} + 4)$, where $j_{\te{max},n}$ is the order of the Taylor expansion of the $n$th momentum eigenvalue.

Collecting terms in \eqref{eq:expansions1}, we have the following expressions for the coefficients of the $\hat{\mathbf{S}}_\lambda$ matrices, for $\lambda = 1$ and $\lambda = n > 1$ respectively
\begin{widetext}
\begin{equation}
    \begin{aligned}
c_{\pm 1,i} &=\sqrt{\pm\omega(1-\omega)}\ps{(\delta_{1 + M-i}\mp \delta_{-1+M-i})-\right.\\&\left.-\sum_{j = 0}^{j_{\te{max},1}}\mat{c}{2j\\j}\f{1}{(j + 1)2^{2j + 1}}\p{\pm\f{\omega - 1}{4\omega}}^{j + 1} \sum_{k = 0}^{2(j + 1)}\mat{c}{2j + 2\\k} (\pm 1)^{k}\p{\delta_{2(j-k) +3 +M- i}\mp \delta_{2(j-k) +1+M - i}}}\,,\\
c_{n,i}&=\sqrt{(n\omega)^2 - 1}\ps{\delta_{i - M} - \sum_{j = 0}^{j_{\te{max},n}}\f{(2j)!}{(j!)^2}\f{2^{-3j - 2}}{(j + 1)}\p{\f{(1-\omega)^2}{1-(n\omega)^2}}^{j+1}\times\right.\\&\left.\sum_{k_{-4} + k_{-2} + k_0 + k_2 + k_4 = j + 1}\f{(j+1)!}{k_{-4}!k_{-2}! k_0! k_2! k_4!}\p{\f12 }^{k_4 + k_{-4}}\p{\f{2 n\omega}{\omega-1}}^{k_2 + k_{-2}} \delta_{4k_4-4k_{-4} + 2k_2 - 2k_{-2} + M - i}}\,.
    \end{aligned}
\end{equation}
\end{widetext}
Finally, we define the matrix of frequencies
\begin{align}
    \mathbf{\Omega} &=\mat{ccccc}{\ddots\\&\omega\\&&0\\&&&-\omega\\&&&&\ddots}\,,
\end{align}
where the even entries are dropped when $V$ has parity. These versions of $\mathbf{\Omega}$ and $\mathbf{S}$ are not to be confused with those used to solve for the physical quasibreather \eqref{eq:LinearMatrix}, as the correct version to use will always be clear from context.

With these definitions, the irrational eigenvalue problem \eqref{eq:fullLinearPert} has been reduced to the polynomial eigenvalue problem
\begin{align}
0&= \sum_{i = 0}^N u^i \mathbf{M}_i\,,
\end{align}
where the matrices $\mathbf{M}_i$ are defined
\begin{widetext}
\begin{align}
\mathbf{M}_i &= -\I \sum_{\lambda = -L}^L c_{\lambda,i} \hat{\mathbf{S}}_\lambda +\ps{2\p{\f{1 - \omega}{2}}^2\mathbf{I} +  \mathbf{\Omega}^2 + \mathbf{L}}\delta_{M - i} + \p{\f{1 - \omega}{2}}^2\mathbf{I}\p{\delta_{M + 4 - i} + \delta_{M - 4 - i}} + (1-\omega)\mathbf{\Omega}\p{\delta_{M + 2-i} + \delta_{M - 2 - i}}\,,\\
\mathbf{L}&={\bm{\nabla}}^2 - \f{\ell(\ell + d - 2)}{r^2} +\f{(d-1)(d-3)}{4r^2}- V_0'' - \f12\sum_{m \in\N_\te{even}}^\infty V_m'' \p{\mathbf{D}_{-m} + \mathbf{D}_{m}} - \f{1}{2\I}\sum_{m \in\N_\te{odd}}^\infty V_m'' \p{\mathbf{D}_{m}-\mathbf{D}_{-m}}\,,
\end{align}
\end{widetext}
where displacement matrices $\mathbf{D}_m$ are the matrices consisting of all 1's on the diagonal of the $m$th block diagonal.
Thus, we have reduced the computation of the Lyapunov exponents to computing the eigenvalues of the generalized eigenvalue problem
\begin{align}
0&=\mat{ccccc}{u\mathbf{I}&-\mathbf{I}\\&u\mathbf{I}&-\mathbf{I}\\&&\ddots&\ddots\\&&&u \mathbf{I}&-\mathbf{I}\\\mathbf{M}_0&\mathbf{M}_1&\dots&\mathbf{M}_{N - 2}&\mathbf{M}_{N-1} + u\mathbf{M}_N}\mat{c}{\delta\theta\\ u\delta\theta\\\vdots\\u^{N-1}\delta\theta}\,.
\end{align}
To summarize, the precision of this approximation can be increased to the desired level by
\begin{enumerate}
    \item increasing the resolution of the radial grid by reducing $\diff r$,
    \item increasing the physical radius of the simulation $r_\te{out}$,
    \item increasing the number $L$ of Floquet blocks kept in the expansion,
    \item increasing the order $j_{\te{max},\lambda}$ of the momentum expansions,
    \item increasing the number of PQB harmonics kept in the background.
\end{enumerate}

\section{\label{sec:simulation}Explicit time evolution --- numerical methods}

Throughout the text, we refer to explicit numerical simulations for validation of our results. Here we outline the numerical setup used to compute the time evolution of the field $\theta$ in the equations of motion \eqref{eq:EOM}, and the methodology used to measure oscillon frequency $\omega$ and radiated power $P$.

The radial equation of motion for the field $\theta(t,r)$ in $3+1$ dimensions is
\begin{align}
    0&=\pd{\theta(t,r)}{t}{2} - \pd{\theta(t,r)}{r}{2} - \f{2}{r}\pd{\theta(t,r)}{r}{} + F(\theta(t,r))\,.
\end{align}
We introduce the variable $v = r\theta$, which eliminates the friction term.
We now discretize time and space, with time steps $\diff t$ and radial steps $\diff r$, and introduce the notation
\begin{align}
    v(N\diff t,M\diff r) = v_N(M)
\end{align}
Finally, we define the ratio $\xi\equiv (\diff t/\diff r)^2$. 
In this notation, the equations of motion lead to the following leading-order finite difference equation
\begin{align}\notag
    v_{N + 1}(M)&= \xi (v_N(M+1)+v_N(M-1))
        + 2(1-\xi) v_N(M) \\&- v_{N-1}(M)
        - (\diff t^2 \diff r M) F(v_N(M)/(\diff r M))\,.
\end{align}
Dirichlet boundary conditions are imposed at the origin by fixing $v_N(0) = 0$. The oscillon is assumed to be evolving in empty space, and therefore the box size must be chosen large enough that the radiation from the oscillon reflected off the outer boundary does not propagate backwards and interfere with the oscillon itself.

The length scale of the $n$th harmonic is $2\pi/\sqrt{(n\omega)^2-m^2}$. During an instance of destructive interference, typically the fifth harmonic dominates, and in rare cases the seventh may contribute significantly. Since $2\pi/\sqrt{7^2 - 1}\approx 0.9$, we choose a safe value of $\diff r = 0.1/m$, about 10 times smaller than the length scale of radiation at the highest possible frequency. We find that $\xi = 1/4$ leads to stable evolution for $\diff r$ of order $0.1/m$. To check that this choice of $\diff r$ is good, we increased the resolution by a factor of $2$ and $4$ which resulted in marginal discrepancies.

The frequency of the oscillon is then measured by tracking the times at which $v_N(1)$ crosses through zero. The outgoing flux is measured outside the oscillon bulk, typically between 20 and 100 in units of the mass. We do not measure the flux too far from the source, since the different frequency modes travel at different velocities, and the PQB formalism does not account for this dispersion.

\bibliography{Bibliography}
\end{document}